\documentclass[11pt, prd, a4paper, showkeys, nofootinbib, twoside, onecolumn]{revtex4-1}
\pdfoutput=1
\usepackage{amsmath,amsfonts,amssymb,amsthm,amssymb,bbm,bm}
\usepackage{pdfsync}
\usepackage{graphicx,import}
\usepackage[latin1]{inputenc}
\usepackage{hyperref}
\usepackage[all]{xy}
\usepackage{stmaryrd}
\usepackage{rotating}
\usepackage{color}  
\usepackage{slashed}
\usepackage{cancel}
\usepackage{array, makecell} %

\numberwithin{equation}{section}

\hypersetup{
    colorlinks,%
    citecolor=blue,%
    filecolor=blue,%
    linkcolor=blue,%
   urlcolor=blue,
   linktoc=page
}

\newcommand{\p}{\partial}

\newcommand{\bit}{\begin{itemize}}
\newcommand{\eit}{\end{itemize}}
\newcommand{\bd}{\begin{description}}
\newcommand{\ed}{\end{description}}

\newcommand{\bc}{\begin{center}}
\newcommand{\ec}{\end{center}}
\newcommand{\Ref}[1]{(\ref{#1})}

\newcommand{\C}{{\mathbb C}}

\newcommand{\SO}{\mathrm{SO}}

\newcommand{\scr}{\scriptstyle}
\newcommand{\sscr}{\scriptscriptstyle\rm}

\newcommand{\be}{\begin{equation}}
\newcommand{\ee}{\end{equation}}
\newcommand{\bea}{\begin{eqnarray}}
\newcommand{\eea}{\end{eqnarray}}
\newcommand{\bs}{\begin{subequations}}
\newcommand{\es}{\end{subequations}}
\newcommand{\nn}{\nonumber}

\newcommand{\w}{\wedge}

\newcommand{\tr}{{\rm Tr}}
\newcommand{\f}{\frac}
\newcommand{\tl}{\tilde}
\def\p{\partial}

\newcommand{\na}{\nabla}

\newcommand{\sd}{\slashed{\delta}}
\newcommand{\sg}{\slashed{\gamma}}

\def\a{\alpha}
\def\b{\beta}
\def\g{\gamma}
\def\d{\delta}
\def\eps{\epsilon}

\def\th{\theta}

\def\k{\kappa}
\def\l{\lambda}
\def\m{\mu}
\def\n{\nu}
\def\x{\xi}
\def\r{\rho}

\def\s{\sigma}
\def\t{\tau}

\def\om{\omega}
\def\G{\Gamma}

\def\Th{\Theta}

\def\Si{\Sigma}

\def\L{\Lambda}
\def\Om{\Omega}

\newcommand{\scri}{\cal I}

\newcommand{\ut}[1]{ \underset{\widetilde{}}{#1}{} }

\newcommand{\fd}{\mathbbm d} 

\DeclareFontFamily{U} {MnSymbolC}{}

\DeclareFontShape{U}{MnSymbolC}{m}{n}{
  <-6> MnSymbolC5
  <6-7> MnSymbolC6
  <7-8> MnSymbolC7
  <8-9> MnSymbolC8
  <9-10> MnSymbolC9
  <10-12> MnSymbolC5 
  <12-> MnSymbolC12}{}
\DeclareFontShape{U}{MnSymbolC}{b}{n}{
  <-6> MnSymbolC-Bold5
  <6-7> MnSymbolC-Bold6
  <7-8> MnSymbolC-Bold7
  <8-9> MnSymbolC-Bold8
  <9-10> MnSymbolC-Bold9
  <10-12> MnSymbolC-Bold10
  <12-> MnSymbolC-Bold12}{}

\DeclareSymbolFont{MnSyC} {U} {MnSymbolC}{m}{n}

\DeclareMathSymbol{\Istar}{\mathrel}{MnSyC}{128}

\newcommand{\FE}[1]{\mathbbm E^{(#1)}}
\newcommand{\eqSi}{\stackrel{\Si}=}

\begin{document}

\title{\Large{Boundary effects in General Relativity with tetrad variables}}

\author{Roberto Oliveri}
\email[email address: ]{roliveri@fzu.cz}
    \affiliation{CEICO, Institute of Physics of the Czech Academy of Sciences, \\Na Slovance 2, 182 21 Praha 8, Czech Republic }
\author{Simone Speziale}
 \email[email address (corresponding author): ]{simone.speziale@cpt.univ-mrs.fr}
    \affiliation{Aix Marseille Univ., Univ. de Toulon, CNRS, CPT, UMR 7332, 13288 Marseille, France}


\begin{abstract}

Varying the gravitational Lagrangian produces a boundary contribution that has various physical applications. It determines the right boundary terms to be added to the action once boundary conditions are specified, and defines the symplectic structure of covariant phase space methods.
We study general boundary variations using tetrads instead of the metric. This choice streamlines many calculations,  especially in the case of null hypersurfaces with arbitrary coordinates, 
where we show that the spin-1 momentum coincides with the rotational 1-form of  isolated horizons.
 The additional gauge symmetry of internal Lorentz transformations leaves however an imprint: the boundary variation differs from the metric one by an exact 3-form.  On the one hand, this difference helps in the variational principle: gluing hypersurfaces to determine the action boundary terms for given boundary conditions is simpler, including the most general case of non-orthogonal corners.
On the other hand, it affects the construction of Hamiltonian surface charges with covariant phase space methods, which end up being generically different from the metric ones, in both first and second-order formalisms. 
This situation is treated in the literature gauge-fixing the tetrad to be adapted to the hypersurface or introducing a fine-tuned internal Lorentz transformation depending non-linearly on the fields. 
We point out and explore the alternative approach of dressing the bare symplectic potential to recover the value of all metric charges, and not just for isometries.
Surface charges can also be constructed using a cohomological prescription: in this case we find that the exact 3-form mismatch plays no role, and tetrad and metric charges are equal. This prescription leads however to different charges whether one uses a first-order or second-order Lagrangian, and only for isometries one recovers the same charges.

\end{abstract}

\maketitle

\tableofcontents

\section{Introduction}

The tetrad description of General Relativity is classically equivalent to the metric one: even though there is an additional gauge symmetry associated with internal Lorentz transformations, the Lagrangian and field equations are equivalent, and physical solutions can be put in one-to-one correspondence. 
There is, however, a subtlety in the presence of boundaries: the \emph{bare} boundary 3-forms that can be read off the arbitrary variation of the tetrad and metric Lagrangians, are \emph{not} equal. The difference is a certain exact 3-form \cite{DePaoli:2018erh}, namely a 2d boundary term, and it is present in both first-order and second-order versions of the Lagrangian. 
In this paper, we revisit and continue the work started in \cite{DePaoli:2018erh} and discuss further implications of the 2d boundary mismatch. 

The variational principle has recently been addressed in metric variables in \cite{Parattu:2015gga,Lehner:2016vdi,Hopfmuller:2016scf}, and in tetrad variables in \cite{Jubb:2016qzt,Wieland:2017zkf}, with a renowned attention to null hypersurfaces and to corner terms. We were particularly motivated by the results of \cite{Lehner:2016vdi}, where a geometric decomposition of the symplectic potential for all types of hypersurfaces was derived, with arbitrary variations in the space-like and time-like cases, and variations restricted to a partial Bondi gauge in the null case; and of \cite{Parattu:2015gga,Hopfmuller:2016scf}, where the null case was studied with arbitrary variations.
These results can be reproduced and elegantly written in terms of differential forms using the tetrad symplectic potential, provided one takes into account the 2d boundary mismatch: 
using the bare tetrad symplectic potential, one reproduces only the 3d bulk part of the metric decomposition, whereas the 2d boundary part is different.
This was shown in \cite{DePaoli:2018erh} for non-null hypersurfaces. In this paper, we complete the analysis for the null case in arbitrary coordinates. These results are presented in Section~\ref{Sec:Boundary}, after an initial Section~\ref{Sec:DPS} that  reviews the mismatch of the bare symplectic potentials.

The  mismatch between tetrad and metric symplectic potentials shows up on the 2d boundary of an individual hypersurface, but cancels out when different hypersurfaces are glued together to form the boundary of a closed region, since it comes from a globally defined exact 3-form. Hence, it does not affect the variational principle. 
There are, nonetheless, two interesting features associated with this mismatch. 
First, the tetrad and metric symplectic potentials give different canonical pairs at the corner: internal vectors instead of spacetime vectors, and a factor of two. This is discussed in Section~\ref{Sec:Pairs}.
Second, gluing together the hypersurfaces turns out to be computationally much simpler with the tetrad potential than with the metric potential, even though the final result is the same. The reason for this lies in the simpler variations of scalar products with internal indices, which use a fixed Minkowski metric. This is discussed in Section~\ref{Sec:Corners}.

The second topic we investigate in this paper is the construction of surface charges with covariant phase space methods \cite{AshtekarReula,WittenCovPhase,LeeWald,Wald:1999wa}. 
In this context, the tetrad-metric mismatch has stronger consequences: two symplectic potentials equal up to an exact form can give different symplectic structures, Hamiltonian generators and charges. This is indeed the case with the bare tetrad and metric symplectic potentials for General Relativity \cite{DePaoli:2018erh}. 
First of all, the bare tetrad potential gives rise to internal Lorentz charges which are absence in metric variables: one is thus associating two inequivalent phase spaces to the same set of physical solutions.  Moreover, the  charges associated with diffeomorphisms obtained from the bare tetrad potential do not coincide with the metric ones. This inequivalence spoils the covariant phase space derivation of the  first law of black hole mechanics from diffeomorphism symmetry. To resolve this problem, it was proposed in \cite{TedMohd,Prabhu:2015vua} to associate the first law with a fine-tuned combination of diffeomorphisms and internal gauge transformations. 

There exist an alternative procedure: one can restore the \emph{full} equivalence of the phase space, and  { as a consequence} of all charges, by dressing the bare tetrad symplectic potential with the exact 3-form of \cite{DePaoli:2018erh}. With such a dressed symplectic potential, the internal Lorentz charges are set to zero, the diffeomorphism charges match the metric ones, and the first law is derived as in the metric theory. 
In other words, the dressed symplectic potential  provides a gauge-invariant phase space for tetrad General Relativity, where by gauge here we mean the internal Lorentz transformations,  {and gives a complementary understanding of the prescription of \cite{TedMohd,Prabhu:2015vua}.}
In Section~\ref{Sec:charges}, we present a comparison of all Noether and Hamiltonian charges using both choices for the symplectic potential, and allowing for arbitrary field-dependent gauge parameters and diffeomorphisms. 
The comparison shows the extent of the mismatch between bare potentials: 
not a single charge in tetrad variables, be it Hamiltonian or Noether, matches the corresponding metric one.
There is one important exception though,  discussed in Section~\ref{Sec:TG}. If one gauge fixes the internal Lorentz symmetry to have a tetrad adapted to the hypersurface, the bare tetrad charges coincide with those of the metric theory for  isometries (but not for general diffeomorphisms). This explains why the mismatch was not observed in \cite{Ashtekar:2008jw}, where the Poincar\'e charges at spatial infinity were recovered using an adapted tetrad.
Since working with an adapted tetrad is quite customary, this may also be the reason why the mismatch went mostly unappreciated so far.

The above considerations are valid whether one uses a second-order Lagrangian or a first-order Lagrangian with independent tetrad and connection variables. In the latter case, one can have an extension of General Relativity to gravity theories with torsion, depending on the matter coupling chosen.
For completeness, we provide in Section~\ref{Sec:Immirzi} explicit formulas in the presence of torsion, and a brief comment on possible new features, in particular in the presence of the Barbero-Immirzi parameter.

One question that arose in doing this work was whether the non-gauge-invariance of the bare phase space in tetrad variables could be dealt with using the cohomological prescription for the charges \cite{Anderson:1996sc,Barnich:2000zw,Barnich:2001jy} (aka Barnich-Brandt (BB) charges), instead of the Hamiltonian prescription of covariant phase space methods. This alternative prescription is free of ambiguities; in particular the freedom to add an exact 3-form to the symplectic potential is eliminated working with the unique weakly-vanishing Noether current. 
This turns out to be the case, as we discuss in Section~\ref{Cohomological method}, even though some care is needed: the order of the Lagrangian now matters. We briefly review the cohomological prescription, and recall two useful facts. First, in General Relativity with metric variables (in the second-order formalism), the BB charges are equivalent to the Hamiltonian ones for isometries, but they differ for general diffeomorphisms by an extra surface term \cite{Barnich:2001jy}. Second, the homotopy operator responsible for the extra surface term has vanishing action for all theories in the first-order formalism. The BB charges for first-order tetrad GR must then always coincide with the Hamiltonian ones with bare tetrad symplectic potential, as indeed found in \cite{Barnich:2016rwk} (see also \cite{Frodden:2017qwh}). Since we have already pointed out that the bare tetrad Hamiltonian charges differ from the metric ones (in both first and second order formalisms), we conclude that also the tetrad BB charges differ from the metric Hamiltonian ones (as well as from the metric BB charges) for general diffeomorphisms. And, in fact, the charges found in \cite{Barnich:2016rwk} do not match the familiar second-order metric ones in general, and the authors follow the same fine-tuning of \cite{TedMohd,Prabhu:2015vua} to recover them in the presence of isometries. 
Having clarified this point, we compute the BB charges with tetrad variables in the second-order formalism. We show that in this case the cohomological prescription leads to vanishing internal Lorentz charges, and all diffeomorphism charges are equal to the metric ones. Therefore the cohomological prescription does solve the non-gauge-invariance problem, but only if one uses a second-order action, and not a first-order one.

This discussion brings to the foreground the fact that the BB prescription leads to different charges depending on whether one uses a first-order or second-order formalism. Tetrad gravity is one example where this difference shows up for BB charges, and we briefly present a second example provided by Yang-Mills theory in first-order and second-order formalism. The equivalence of the charges is obtained only for isometries, namely Killing diffeomorphisms in General Relativity and parallel gauge transformations in Yang-Mills theory.
Yet, surface charges can have applications beyond the case of isometries or asymptotic symmetries, hence this difference is worth pointing out in our opinion. 
This is in contrast to the Hamiltonian prescription for the charges, where the change from second-order to first-order is harmless for any gauge transformation and not just for isometries (unless one looks at modified theories of gravity with torsion).

Throughout the paper we fix units $16\pi G=1$.
We use signature with mostly plus; Greek letters for spacetime indices and capital Latin letters for internal indices. The spacetime Hodge dual is denoted by $\star$. To simplify the notation for the pull-back of differential forms, we will use $\eqSi$ to refer to equalities valid for the pull-back of $3$-forms on the hypersurface $\Si$. It will be implicitly assumed that subsequent equalities in the same equation are also pulled-back.
%

\section{Variation of the gravitational action with boundaries} \label{Sec:DPS}

The tetrad Lagrangian for General Relativity  can be elegantly written as a $4$-form,
\be\label{SEC}
L_{\scr e}  = 
\f12\eps_{IJKL}~e^I\w e^J \w \Big(F^{KL} - \f\L6 ~ e^K\w e^L\Big).
\ee
Here, $F^{IJ}= d \om^{IJ} +\om^{IK}\w\om_K{}^J$ is the curvature 2-form, $\om^{IJ}\equiv \om^{IJ}(e)$ is the Levi-Civita Lorentz connection, and $\Lambda$ is the cosmological constant. 
The associated action principle must be supplemented by appropriate 3d and 2d boundary terms, depending on the boundary conditions chosen for the dynamical fields.
We will come back to these in Section~\ref{Sec:Corners} below.

Varying the Lagrangian \Ref{SEC}, we obtain the field equations and an exact 4-form, induced by integrating by parts the identity $\d F(\om)=d_\om(\d\om)$,
\begin{align}\label{dL}
\d L_{\scr e} &= \d e^I\w \FE{e}_I  + d\th_{\scr e}(\d),
\end{align}
where
\begin{subequations} 
\begin{align}
 \FE{e}_I &= \eps_{IJKL} ~e^J\w \left(F^{KL}-\f23\L ~e^K\w e^L \right) \label{FE1},\\
\th_{\scr e}(\d) &= \f12\eps_{IJKL}~e^I\w e^J\w \d\om^{KL} \label{thEC1}.
\end{align}
\end{subequations}
In terms of Hodge duals, which will be used below,
\begin{subequations} 
\begin{align}\label{EEe}
\left(\star \FE{e}_I\right)^\m &= -2\left(G^\m_I+\L e^\m_I\right)
\\\label{sthEC}
\left(\star\th_{\scr e}\right)^\m &= \f1{3!}\left(\th_{\scr e}\right)_{\n\r\s}\eps^{\n\r\s\m} = 2e^{[\m}_I e^{\n]}_J\d\om_\n^{IJ}.
\end{align}
\end{subequations}
In the above formulas, we always intend for $\om^{IJ}$ a Levi-Civita connection; hence $\d\om^{IJ}$ is a short-hand notation for the full expression in terms of tetrad variations, similarly to the notation used in the metric formalism. 

We want to compare the  tetrad boundary variation \eqref{thEC1}-\eqref{sthEC} with the corresponding one in metric variables.
The tetrad is related to the spacetime metric by $g_{\m\n}=e^I_\m e^J_\n \eta_{IJ}$, and the Lorentz connection to the tangent bundle connection by  $\G^\m_{\n\r}= e^{\m}_I D_\n e^I_{\r} = e^{\m}_I\left(\p_{\n}e^I_\r + \om^{IJ}_\n e_{J\r}\right)$, or equivalently, by
\be \label{defom}
{\om}^{KL}_\m=e_\n^K\na_\m e^{\n L}.
\ee 
From this, it follows that $F^{IJ}_{\m\n}=e^{I\r}e^{J\s}R_{\r\s\m\n}$, and \Ref{SEC} is equivalent to the Einstein-Hilbert (EH) Lagrangian,
\be\label{LEH}
L_{\scr g} = \left(g^{\m\n}R_{\m\n} -2\L \right)\eps,
\ee
where $\eps$ is the volume 4-form. 
The variation of the EH Lagrangian gives the field equations plus an exact 4-form, induced by the identity $g^{\m\n}\d R_{\m\n}= 2\na_\m (g^{\r[\s} \d \G^{\m]}_{\r\s})$,
\be
\d L_{\scr g} = \left(G_{\m\n}+\L g_{\m\n}\right)\d g^{\m\n}\eps+ d\th_{\scr g}(\d),
\ee
where 
\begin{subequations}
\begin{align}
\th_{\scr g}(\d) &=\f1{3!} (\star\th_{\scr g})^\m \eps_{\m\n\r\s}~dx^\n\w dx^\r \w dx^\s , \\
\left(\star\th_{\scr g}\right)^\m &= 2 g^{\r[\s} \d \G^{\m]}_{\r\s} = 2g^{\m[\r}g^{\n]\s} \na_\n\d g_{\r\s}. \label{thg}
\end{align}
\end{subequations}

The potentials $\th(\d)$ for the boundary variations are defined up to the addition of an exact 3-form, and hereafter we will refer to \Ref{sthEC} and \Ref{thg} as \emph{bare} potentials. They play an important role in many physical applications, notably in covariant phase space methods: they are used to define the \emph{symplectic potential} $\Th(\d)$ associated with a complete or partial Cauchy hypersurface $\Si$, 
\be \label{Thdef}
\Th(\d) :=\int_{\Si}\theta(\d).
\ee
With a slight abuse of language, we will also refer to the 3-form integrands as symplectic potentials.

\subsection{Matching the symplectic potentials: the dressing 2-form}

The Lagrangians in tetrad and metric variables are equivalent, and so are the variations and the field equations. There is, though, no guarantee that the bare symplectic potentials are equivalent, because of the cohomology ambiguity in extracting the potential from an exact form. 
To compare the bare tetrad and metric potentials, we take the functional variation of the Lorentz connection \Ref{defom} and plug it in~\eqref{sthEC}. This leads to
\begin{align}
\left(\star\th_{\scr e}\right)^{\mu} &= 2e^{[\m}_I e^{\n]}_J \left[ \d e^I_\l \nabla_{\n}e^{J\l}  + e^I_\l \d \G^{\l}_{\n \rho} e^{J\rho}  
+ e^I_\l \na_\n\d e^{J\l} \right] \nn\\
&= 2\d e^{[\m}_I \na_{\n}e^{\n]I} + 2 g^{\r[\n} \d \G^{\m]}_{\n\r} + 2e^{[\n}_I \na_{\n} \d e^{\m]I}\nn\\ 
&= \left(\star\th_{\scr g}\right)^{\mu} + \na_{\n}\left(2e^{[\n}_I  \d e^{\m]I}\right),
\end{align}
exposing explicitly their non-equivalence.
Taking the Hodge dual of this expression (see Appendix~\ref{AppT1} for explicit formulas), we can equivalently write it as
\be\label{thda}
\th_{\scr g}(\d) = \th_{\scr e}(\d) + d\a(\d),
\ee
where
\be \label{DPS}
\a(\d) = \star (e_{I}\w \d e^{I}) = -\frac{1}{2}\eps_{IJKL}~e^I\w e^J \left(e^{\r K}\d e^L_\r\right)
\ee
is the 2-form  introduced in \cite{DePaoli:2018erh} (and shortly after in \cite{Dolan:2018hfo,Gomes:2018shn}; see also \cite{Bodendorfer:2013jba} for a related expression in the canonical formalism), and hereafter dubbed DPS 2-form or dressing 2-form.\footnote{Its original derivation \cite{DePaoli:2018erh}  was in the first-order formalism, and included the contribution of the Barbero-Immirzi term; see Section~\ref{Sec:Immirzi} below. }

Eq.~\eqref{DPS} shows that the bare tetrad and metric potentials differ by an exact 3-form. The origin of this difference lies in the additional (gauge) structure the tetrad field: the variation $e_{I}\w \d e^{I}$ has no metric equivalent, since it corresponds to the antisymmetric part of the tensorial tetrad perturbations.

\section{Geometric decomposition of the boundary variation} \label{Sec:Boundary}

The first application of the boundary variations that we consider is the variational principle.
To identify the boundary terms needed for a well-defined variational principle with given boundary conditions, it is useful to separate the variations in tangential and orthogonal pieces, and give them an interpretation in terms of intrinsic and extrinsic geometry of the hypersurface. We show in this Section how this familiar procedure in metric variables (see, \emph{e.g.}, \cite{Parattu:2015gga,Lehner:2016vdi,Hopfmuller:2016scf} for recent work) can be performed using tetrad variables, and comment on the role of the mismatch \Ref{thda}.
We specify the hypersurface with a Cartesian equation, and keep it fixed while allowing arbitrary variations of the metric and tetrad fields. 
For restricted variations preserving the induced metric on the hypersurface with adapted tetrads, see \cite{Jubb:2016qzt}.

\subsection{Non-null hypersurfaces}\label{SecNN}

The bulk part of the potential has a familiar form in metric variables. The use of tetrads gives a different derivation thereof, but the only novelty to focus on here is really the 2d boundary mismatch.

\subsubsection*{Preliminaries}

Consider an  hypersurface $\Si$ with normal 1-form $n_\m$. If $\Si$ is either space-like or time-like, we can normalize $n_\m$ to be unit-norm, and denote $s:=n^2=\mp1$.
We define the projector tensor  $q_{\m\n} := g_{\m\n}-sn_\m n_\n$, and the extrinsic curvature of the hypersurface by $K_{\m\n}:=q^\r_{\;\m} q^\s_{\;\n} \na_\r n_\s$, with trace $K=\na_\m n^\m$.  
The extrinsic curvature is a symmetric tensor, and we can write its tetrad projection as $K^\m_{\;I} =q^\r_{\;I} \na_\r n^\m$. 
We denote the volume 3-form by $d\Si := i_n \eps =s n^{\m}d\Si_{\m}$, where $d\Si_\m:=sn_\m d\Si$ is the  oriented volume element in the conventions  of \cite{Lehner:2016vdi}. Accordingly, the pull-back of a 3-form $\th$ is
\begin{align}\label{3dpb}
\th \eqSi  \f s{3!}\th_{\m\n\r} \eps^{\m\n\r\s} n_\s d\Si = \f1{3!}\th_{\m\n\r} \eps^{\m\n\r\s} d\Si_\s = (\star\th)^\m d\Si_\m.
\end{align}
If $\Phi(x)=0$ denotes the Cartesian equation of the hypersurface, we can take adapted coordinates $x^\m=(\Phi,y^a)$ and then $q_{ab}=g_{ab}$, $q:=\det q_{ab}>0$ and $d\Si=\sqrt{-sq}~d^3y$.

If the hypersurface has a boundary $\p\Si$, we denote by $\hat r$ its outgoing unit-norm normal within $T\Si$, so that $\hat r_\m n^\m=0$. We will restrict our attention to the case when $\p\Si$ is space-like, thus $\hat r^2=-s$ and we can write the projector on $\p\Si$ as
$\g_{\m\n}=q_{\mu\nu} + s \hat r_\m\hat r_\n = g_{\m\n}-sn_\m n_\n +s\hat r_\m\hat r_\n$. 
We denote the area 2-form by $dS:=i_{\hat r}d\Si = n^{\rho}\hat r^{\s}dS_{\r \s}$,  where $dS_{\r \s} =- 2n_{[\rho}\hat r_{\s]}dS$ is the oriented surface element with both outgoing normals. 
Accordingly, the pull-back of a 2-form $\a$ on $\p\Si$   is
\begin{align}
\a &\stackrel{\p\Si}= \f 1{2}\a_{\m\n} \eps^{\m\n\r\s} n_\r \hat r_\s dS = -\f1{4}\a_{\m\n} \eps^{\m\n\r\s} dS_{\r\s} = 
-\f12 (\star\a)^{\m\n} dS_{\m\n}.
\end{align}
If $R(y)=0$ denotes the Cartesian equation of the boundary of the hypersurface, we can take adapted coordinates $x^\m=(\Phi,R,\th^A)$ and then $\g_{AB}=g_{AB}$, $\g:=\det \g_{AB}$ and $dS=\sqrt{\g}~d^2\th$.

\subsubsection*{Geometric decomposition}

The geometric decomposition of the bare symplectic potential \Ref{thEC1} in tetrad variables for arbitrary variations was discussed in \cite{DePaoli:2018erh}. 
We give here the result, and review its derivation below.
Using the short-hand notation $\Si^{IJ} = e^{I}\w e^J$, one gets
\begin{subequations}\label{The1}\begin{align} 
\Th_{\scr e}(\d) =&s \int_\Sigma \eps_{IJKL}\big[\d \Si^{IJ} \w n^{K} d_\om n^L -\d \big(\Si^{IJ} \w n^{K} d_\om n^L \big)  \big] + 
s\int_{\p\Si}\eps_{IJKL} ~\Si^{IJ} n^K \d n^L \label{ThEC} \\
\hspace{2mm} = &s\int_\Si \left(K_{\m\n} \d q^{\m\n} -2\d K \right) d\Si
- \int_{\p\Si} 2\hat r_I  \d n^I dS. \label{maggica1}
\end{align}\end{subequations}
The bulk term can be recognized as the familiar EH result; see, \emph{e.g.}, \cite{BurnettWald1990,Lehner:2016vdi}. The integrand of the boundary term is instead different, as to be expected from \Ref{thda}. It can be rewritten in terms of spacetime vectors as follows,
\be\label{silva}
2 \hat r_I  \d n^I = 2\hat r_\m  \d n^\m +2\hat r_I  n^\m \d e^I_\m =\hat r_\m  \d n^\m + (n^\m \hat r_{I}  - \hat r^\m n_I)\d e_\m^I.
\ee
In the last step we used $\hat r_\m\d n^\m = - \d\hat r_\m n^\m = -(\hat r_{I} n^\m + n_I\hat r^\m)\d e^I_\m - n_\m \d \hat r^\m$, and 
$n_\m \d \hat r^\m=0$ since $\hat r^\m$ has no components outside $T\Si$.

On the other hand, the pull-back to the boundary $\p\Si$ of $\a(\d)$, see Eq.~\eqref{DPS}, gives
\begin{align}\label{DeRossi}
\int_{\p\Si}\a(\d) &=-\f12\int_{\p\Si} \eps_{IJKL}~\Si^{IJ} e^{\r K}\d e^L_\r 
=\int_{\p\Si} \left(n^\m \hat r_{I}  - \hat r^\m n_{I}\right)\d e^I_\m dS.
\end{align}
Adding \Ref{DeRossi} to the bare tetrad potential \Ref{maggica1}, we see that this contribution cancels the round-bracket term proportional to the variation of the tetrad in \Ref{silva}, giving
\begin{align}\label{thEH1}
\Th_{\scr g}(\d) = \Th_{\scr e}(\d)+\int_{\p\Si}\a(\d)
= s\int_\Si \left( K_{\m\n} \d q^{\m\n}-2\d K \right) d\Si - \int_{\p\Si} \hat r_\m  \d n^\m dS,
\end{align}
which is the complete metric result including the boundary term.
The difference in the boundary terms between the bare and dressed potentials is in a factor of 2 and an additional variation of the tetrad field.

The formula \Ref{The1} provides an independent and, to our taste, shorter and more elegant way of deriving the geometric expression of the gravitational symplectic potential.
Given its relevance, we review below the proof done in
\cite{DePaoli:2018erh}.

\subsubsection*{Proof of Eq.~\eqref{The1}} \label{PnDNN}
To derive the geometric decomposition \eqref{The1}, we start from an
expression for the trace of the extrinsic curvature in tetrad variables:
\begin{align} \label{cusa1}
\f12\eps_{IJKL}~e^I\w e^J \w n^K d_\om n^L &\stackrel{\Si} = \f 12\eps_{IJKL} e^I_\m e^J_\n n^K D_\r n^L s \eps^{\m\n\r\s} n_\s d\Si \nn\\
& = s \left (n^2\na_\m n^\m - n_I n^\m D_\m n^I \right) d\Si\nn\\ 
&= Kd\Si,
\end{align}
where we used the inverse tetrad identity \Ref{ee} and the fact that $n_I D_\m n^I=0$.
Comparing the variations of the first and last expressions will allow us to derive \Ref{The1}.
The variation of the first expression gives
\begin{align} \label{cusa2}
\f12\eps_{IJKL}~\d\big( e^I\w e^J \w n^K d_\om n^L\big) &= \f12\eps_{IJKL}\big[\Si^{IJ}\w n^K\d\om^{LM} n_M +\d\Si^{IJ}\w n^Kd_\om n^L 
\nn\\& \hspace{-2cm} +2 \Si^{IJ}\w \d n^K d_\om n^L - d_\om \Si^{IJ} n^K\d n^L\big] + \f12 d\left(\eps_{IJKL}~\Si^{IJ} n^K\d n^L\right),
\end{align}
after an integration by parts to remove the variations of derivatives. 
The first term in the right-hand side contains the symplectic potential, since its pull-back reads as
\begin{align}
\f12\eps_{IJKL}~ \Si^{IJ}\w n^K\d\om^{LM} n_M \stackrel{\Si}= - n_I e^\m_J \d\om^{IJ}_\m d\Si \stackrel{\Si}= -\f s2 \th_{\scr EC}(\d).
\end{align}
The second term gives
\begin{align}
\eps_{IJKL}~\d e^I\w e^J \w n^K d_\om n^L &\stackrel{\Si}  
=- \d e^I_\m \left(q^\r_I \na_\r n^\m - q^\m_I \na_\r n^\r \right)d\Si \nn\\ \label{cusa3}&= \f12 \left(K_{\m\n} - K q_{\m\n}\right)\d q^{\m\n} d\Si, 
\end{align}
where we used the inverse tetrad identity \Ref{eee} in the first equality, and the identities 
\begin{subequations}
\begin{align}
& K_{\m\n}\d q^{\m\n} = q^\r_\m q^\n_\s \na_\s n_\r \d g^{\m\n} = -2 K^\m_I \d e_\m^I, \\
& \d q = - q q_{\m\n}\d q^{\m\n} = - q q_{\m\n}\d g^{\m\n} = 2q q^\m_I \d e_\m^I
\end{align}
\end{subequations}
in the second equality.
The third term vanishes identically because $n$ is unit-norm, and the fourth term vanishes because the Levi-Civita connection is torsion-free. We are left with the boundary term whose pull-back, with our orientation conventions, gives
\begin{align}
& \f12\eps_{IJKL}~ e^I\w e^J n^K\d n^L \stackrel{\p\Si}= \f12\eps_{IJKL}~ e^I_\m e^J_\n n^K\d n^L \eps^{\m\n\r\s} n_\r \hat r_\s dS
= - s \hat r_I\d n^I dS. 
\end{align}

Coming back to Eq.~\Ref{cusa1}, the variation of the final expression gives 
\be\label{deltaKS}
\d(Kd\Si) = \Big(\d K -\f12 K q_{\m\n}\d q^{\m\n}\Big) d\Si.
\ee
Putting everything together, 
the second term of \Ref{deltaKS} cancels with the second term of Eq.~\Ref{cusa3}, and we find
\begin{align}\label{cusa4}
\int_\Si \d K d\Si =  -\f s2 \Th_{\scr EC}(\d) 
+ \f12 \int_\Si K_{\m\n} \d q^{\m\n}d\Si - s\int_{\p\Si} \hat r_I\d n^I dS.
\end{align}
Isolating the symplectic potential, we arrive at Eq.~\eqref{maggica1}.

\subsection{Null hypersurfaces} \label{SecN}

This case requires a longer discussion, because  the bulk part of the potential presents some subtleties already in metric variables, mainly due to the lack of a standard normalization of the hypersurface normal. We use the fact that picking a foliation of the null hypersurface allows us to introduce a preferred normalization, and bridge among various results in the literature. 
We then discuss the boundary mismatch using tetrads, which is essentially the same as in the non-null case.

\subsubsection*{Preliminaries}

To distinguish the case of a null hypersurface, we will refer to it as $\cal N$. Its normal 1-form $n_\m$ has vanishing norm, $n^2=0$, 
and this also means that $n^\m$ is tangent to $\cal N$. Being hypersurface orthogonal and null at $\cal N$, it is automatically geodesic, namely $n^\m\na_\m n_\n = k_{(n)}n_\n$, with $k_{(n)}$ the inaffinity or  tangential acceleration. 
A null hypersurface is thus always ruled by null geodesics. The pull-back of the metric is degenerate, with signature $(0,+,+)$, and  has 5 components. These define a complex 3d dyad up to an SO(2) rotation,
\be\label{induceddyad}
g_{\m\n} \stackrel{\cal N}{=} m_{(\m} \bar m_{\n)}.
\ee
The equivalence class of $n^\m$ up to rescalings can be defined intrinsically to $\cal N$ as the null eigenvector of the induced metric, and  $\eps^{(2)}:=i m \w\bar m$ provides an area 2-form.
If we define the hypersurface with Cartesian equation $\Phi(x)=0$, the normal will generically be of the form
\be\label{nnull}
n_\m = -f\p_\m \Phi,
\ee
with $f>0$ so to have the vector future-pointing. 
In adapted coordinates $(v=\Phi,y^a)$, with $a=1,2,3$, the volume 3-form on $\cal N$ is 
\be\label{dN}
d{\cal N}:= \f{\sqrt{-g}}f d^3y,
\ee
and  $d{\cal N_{\mu}} = - n_{\m}d{\cal N}$
is the oriented volume element. Accordingly, the pull-back of a 3-form on $\cal N$ is
\begin{align}\label{3dpbn}
\th \stackrel{\cal N}= - \f 1{3!}\th_{\m\n\r} \eps^{\m\n\r\s} n_\s d{\cal N} = \f1{3!}\th_{\m\n\r} \eps^{\m\n\r\s} d{\cal N}_\s = (\star\th)^\m d{\cal N}_\m.
\end{align}

Since the vector is null, there is no preferred normalization and thus no preferred choice of $f$. The choice $f=1$ is often made to simplify some calculations. Another convenient choice becomes available if one picks a foliation of $\cal N$ given by the level sets of a parameter  $r$ along the null geodesics (not necessarily affine). We can then require that $n^\mu$ transports slices into slices, namely $i_n dr =1$. This fixes $f=-1/g^{vr}$, and $n^{\mu} = (\partial_r)^{\mu} +b_1^{\mu}$, where $b_1^{\mu}$ is the shift vector of the 2+1 foliation. 
This choice is `canonical' in the following sense. Denote $\th^A$,  $A=2,3$, the coordinates on the
2d slices $S$ of the foliation, with induced space-like metric $\g_{AB}$, with determinant $\g>0$, and pull-back of the area 2-form $\eps^{(2)}\stackrel{S}=dS:=\sqrt{\g}d^2\th$.
Since the inverse spacetime metric has vanishing component $g^{vv}=0$, an explicit calculation shows that
\be
\sqrt{-g} = N \sqrt{\g},
\ee 
where $N:=-1/g^{vr}$ is identified with the lapse function. Therefore the normalization
\be\label{f=N}
f=N = \f{\sqrt{-g}}{\sqrt{\g}} \quad \Leftrightarrow \quad i_n dr=1
\ee
is `canonical' in analogy with the $f=N$ normalization occurring  in the  ADM $3+1$ decomposition. 
And indeed, in the canonical analysis on null foliations, it is  \Ref{nnull} with $f=N$ that appears in the momenta conjugated to the induced metric on a null hypersurface. More on this below. 

With the choice \Ref{f=N}, 
the volume 3-form simplifies to $d{\cal N}=dr dS$. This volume form is foliation dependent. However, we can require $i_n dr=1$ for any foliation of the null generators, and $d{\cal N}=dr dS$ always, if we restrict the $\th^A$ coordinates to be constant along the generators, namely the shift vector $b_1^\m$ to vanish.\footnote{We thank Jos\'e Luis Jaramillo for clarifying this point to us.} See \cite{Gourgoulhon:2005ng} for more details.

In working with null hypersurfaces, it is very convenient to use the Newman-Penrose (NP) formalism.
This is easiest done if $\cal N$ is part of a null foliation, so that $n_\m$ is null everywhere. 
In this case we can pick any transverse null vector $l^\m$ such that $n_{\mu}l^{\mu}=-1$, and define the spacetime projector 
\be\label{gamma}
\g_{\m\n} = g_{\m\n} +l_\m n_\n + n_\m l_\n = 2m_{(\m}\bar m_{\n)}.
\ee 
The forms $(m_\m,\bar m_\m)$ provide an extension of the complex dyad \Ref{induceddyad} off $\cal N$, and the set $(l,n,m,\bar m)$ forms a NP tetrad.
To fix ideas, we will think of $\cal N$ as a section of a past light-cone.\footnote{For a future light cone, it is convenient to denote the normal by $l_\m$, to match with the NP literature.} 
If  
$n^2\neq 0$ off $\cal N$, we can  identify $n_\m$ as one element of a NP tetrad   only at $\cal N$. Most of the formulas below are local on $\cal N$ and will still apply, but in some cases there are additional contributions from the non-zero $\p_\m n^2\propto n_\m$. We will procede assuming $n^2=0$ everywhere, and point out at the relevant places  the additional contributions.

With the projector \Ref{gamma}, one builds the null-hypersurface analogue of the extrinsic geometry, 
namely the deformation tensor and its decomposition into shear and expansion,
\be
B_{\m\n}:=\g^\r_{\;\m} \g^\s_{\;\n} \na_\s n_\r = \s_{(n)\m\n} +\f12\g_{\m\n}\th_{(n)}.
\ee
The divergence of the normal, which gives the trace of the extrinsic curvature in the non-null case, now gives
\be\label{tk}
\na_\m n^\m = \th_{(n)} + k_{(n)}.
\ee
The geodesy of $n^\m$  guarantees that the expansion $\th_{(n)}$ and the projected shear $\l:=\bar m^\m \bar m^\n \s_{(n)\m\n}$ are independent of the choice of $l^\m$. 
We recall that the freedom in picking $l^\m$ is a 2-parameter family of Lorentz transformations preserving $n^\m$,
\be\label{classII}
l^\m\mapsto l^\m+\bar b m^\m +b \bar m^\m+|b|^2 n^\m,\qquad m^\m\mapsto m^\m+b n^\m, \qquad b\in\C.
\ee
If a foliation of $\cal N$ is chosen, the freedom \Ref{classII} can be used to adapt the NP tetrad to the foliation, by making $(m^\m,\bar m^\m)$ integrable and tangent to cross sections of the geodesics.

If $\cal N$ has a space-like boundary $\p \cal N$, we can always take a foliation such that the boundary is part of it, namely it is defined by a level set of $r$. We can then adapt the NP tetrad to this foliation that contains the boundary.
With our choices of orientations and $\cal N$ a past light-cone, the pull-back of a 2-form on the boundary is
\be\label{2dnpb}
\a \stackrel{S} = \f 1{2}\a_{\m\n} \eps^{\m\n\r\s} l_\r n_\s dS.
\ee
Choosing a foliation of $\cal N$ misaligned with the boundary will introduce an additional corner diffeomorphism \cite{Reisenberger:2007ku,Reisenberger:2018xkn,Hopfmuller:2016scf} that will be mentioned below.

\subsubsection*{Variations}

An important aspect of working with a null hypersurface is that it automatically imposes a partial gauge fixing on the metric. This can be made manifest in adapted coordinates with $v=\Phi$, where  $n^2=0$ translates to $g^{vv}=0$ at $\cal N$. Therefore, one \emph{cannot} consider arbitrary variations of the metric tensor if one wants to restrict attention to a null hypersurface, but only those with $\d g^{vv}=0$.\footnote{This fact has a subtle consequence in the canonical formalism based on null foliations \cite{dInverno:1980kaa,Torre:1985rw,Goldberg:1992st, IoSergeyNull}: the partially gauge fixed action depends on 9 metric components only, and one loses at first sight one of Einstein's equations. This ``missing'' equation is usually recovered through a somewhat ad-hoc extension of the phase space. The situation is improved working in the first order formalism, where all equations are recovered without extensions of the phase space; see \cite{IoSergeyNull}.} 
In covariant terms, the allowed variations are restricted to those satisfying
\be\label{ndn}
n_\m\d n^\m=0.
\ee
This follows from $\d n_\m =(\d\ln f) n_\m$, allowing here for the most general set-up where $f$ can depend on the metric,
which in turns implies $n^\m\d n_\m=0$ and \Ref{ndn}.
A metric-dependent normalization also contributes to the variation of inaffinity, since 
\be\label{dkn}
\d k_{(n)} = n^\m \na_\m \d \ln f + \d n^\m \na_\m\ln f.
\ee 
If $\cal N$ is an individual null hypersurface in the foliation, the above variation has an additional contribution
$
\tfrac12 (\d\ln f) \,   l^\m\p_\m n^2.
$

Since the null hypersurface is  ruled by its geodesics, it is often convenient to pick coordinates $(r,\th^A)$ where $r$ is a parameter along the geodesics, not necessarily affine, and $\th^A$ are constant along the geodesics. This introduces two additional gauge-fixing conditions $g^{vA}=0$, and we will refer to it as partial Bondi gauge for historical reasons \cite{BMS}. 
In the partial Bondi gauge only the component $n^r$ is allowed to vary, since 
$n^\m = -N g^{v\m} = -Ng^{vr} \d^\m_r$.
If we then choose the `canonical' normalization \Ref{f=N}, all components are fixed and $\d n^\m=0$. 
Notice that this is achieved while leaving one last coordinate freedom, reparametrizations of $r$. In particular the inaffinity is still arbitrary and so its variation, given now by just the first term of \Ref{dkn}.
This partial gauge-fixing and `canonical' normalization is the set-up used in \cite{Lehner:2016vdi}.
The complete Bondi gauge corresponds to fixing the remaining gauge freedom in terms of the $r$ coordinate, requiring it for instance to be an area coordinate, or an affine parameter for the null geodesics. The latter condition can be imposed fixing $g^{vr}=-1$ as is typically done in the NP analysis of gravitational radiation. This implies a unit lapse, $N=1$, and \Ref{dkn} vanishes in a greement with the fiex affinity of $r$.

Another useful relation is the variation of the normal $n^\m$ along the transverse vector $l^\m$, which is given by
\be\label{ldn}
l_\m \d n^\m = l^\m\d n_\m +\d \ln\sqrt{-g}-\d\ln\sqrt{\sg} =\d \ln \left(\frac{\sqrt{-g}}{f \sqrt{\sg}} \right),
\ee
where we introduced the notation $\d\ln{\sg}:=- \g_{\m\n}\d\g^{\m\n}$. Notice that the quantity $\sg$ only makes sense inside a variation, since \Ref{gamma} has zero determinant. If we fix a foliation and adapt the NP tetrad to it, then $\d\ln{\sg}=\d\ln \g$. 
It follows that in the `canonical' normalization \Ref{f=N}, we have
\be
l_\m\d n^\m\stackrel{f=N}=0.
\ee
With a non-adapted NP tetrad instead, $\d\ln{\sg}\neq \d\ln \g$. The difference is a variation of the 2d shift vector of the foliation of the null hypersurface.
Therefore these two variations will again coincide if the shift vector vanishes, namely in the partial Bondi gauge $g^{vA}=0$. 

We have thus highlighted two special properties of the partial Bondi gauge: it makes it possible to use $\sqrt{\g}drd^2\th$ as null volume form independently of the choice of foliation, and it makes its logarithmic variation equal to $-(1/2)\g_{\m\n}\d\g^{\m\n}$, independently of the choice of $l_\m$.

\subsubsection*{Geometric decomposition}

After these preliminaries, we can now state the main new result of this Section, namely the geometric decomposition of the tetrad symplectic potential on a null hypersurface $\cal N$. We first give the result and discuss it, and provide a detailed derivation below. The decomposition reads as
\begin{subequations}\label{thECnull2}
\begin{align}\label{thECnull}
\Th_{\scr e}(\d) =&\int_{\cal N}\eps_{IJKL} \big[ \d \Sigma^{IJ} \w l^K d_\om n^L 
-\d\big(\Si^{IJ} \w l^{K}d_{\omega}n^L \big) + \Sigma^{IJ}\w \left(\d l^{K}d_\om n^L + \d n^{K} d_{\om}l^L \right)\big] \nn\\
&  + \int_{\p \cal N}\eps_{IJKL}\Sigma^{IJ} l^{K} \d n^L
\\\label{thECnull1}
=& \int_{\cal N}\big[ -B_{\m\n} \d \g^{\m\n} + 2 \d \left(\th_{(n)} + k_{(n)} \right) + 2  \om_{(n)\m}\d n^\m \big]d{\cal N}
+ \int_{\p\cal N} 2 l^I \d n_I dS,
\end{align}
\end{subequations}
where
\be\label{defomega}
\om_{(n)\m}:=\eta_\m + k_{(n)}  l_\m,\qquad 
\eta_\m:=\g^\r_{\;\m} l^\s \na_\r n_\s,
\ee
and we recall that the volume form depends on the normalization of $n_\m$ as in \Ref{dN}.
It can be explicitly checked that the bulk term is independent of the choice of $l_\m$, namely invariant under \Ref{classII}. We have chosen it here so that it acts as normal of the boundary $\p\cal N$. The 2d boundary term can be rewritten in terms of spacetime vectors as follows,
\be\label{statecollege}
2l_I\d n^I = l^\m\d n_\m +l_\m\d n^\m + \left(n^\m l_I-l^\m n_I\right)\d e^I_\m.
\ee

To derive the geometric decomposition for the symplectic potential in metric variables, we evaluate the pull-back of the 2-form $\a(\d)$,
\begin{align}
\int_{\p\cal N} \a(\d) &=-\f12\int_{\p\cal N} \eps_{IJKL}e^I\w e^J e^{\r K}\d e^L_\r 
=\int_{\p\cal N} \left( l^\m n_{I}  - n^\m l_{I}\right)\d e^I_\m dS.
\end{align}
Adding this contribution to \Ref{thECnull1}, we obtain a formula for the dressed, metric-equivalent symplectic potential:
\begin{align}\label{thEHn} 
\Th_{\scr g}(\d) =& \int_{\cal N} \big[  -B_{\m\n} \d \g^{\m\n} + 2 \d ( \th_{(n)} + k_{(n)} ) +2 \om_{(n)\m}  
\d n^\m \big] d{\cal N} \nn\\ 
& + \int_{\p\cal N} \left(l^\m\d n_\m +l_\m \d n^\m\right) dS.
\end{align}

Let us briefly compare this expression with the ones already present in the literature.
If we introduce a foliation of $\cal N$, we can take the canonical normalization $f=N$ and adapt the NP tetrad to it. Then,
\begin{align}\label{thECnullfol}
\Th^{\sscr fol}_{\scr g}(\d)
=& \int_{\cal N}\big[ -B_{\m\n} \d \g^{\m\n} + 2 \d \left(\th_{(n)} + k_{(n)} \right) + 2  \eta_{(n)\m}\d n^\m \big]{\sqrt \g}dr d^2\th
+ \int_{\p\cal N} l^\m\d n_\m dS.
\end{align}
The bulk term coincides with the one of \cite{Hopfmuller:2016scf}. The boundary is only one of  two terms found in \cite{Hopfmuller:2016scf}. 
Notice, however, that we assumed the boundary to be part of the foliation (or in other words, we chose a foliation adapted to the boundary), whereas it appears to be completely general in \cite{Hopfmuller:2016scf}. We leave this question open for future work.

If we furthermore restrict the variations to those with $\d n^\m=0$, \Ref{thEHn} reduces to 
\begin{align}\label{thEHnB} 
\Th^{\sscr Bondi}_{\scr g}(\d) &= \int_{\cal N} \left[  -B_{\m\n} \d \g^{\m\n} + 2 \d \left( \th_{(n)} + k_{(n)}\right) \right] {\sqrt \g}dr d^2\th
 + \int_{\p\cal N} l^\m\d n_\m  dS.
\end{align}
This matches in both bulk and boundary terms the expression given in \cite{Lehner:2016vdi} in the same set-up with a partial Bondi gauge. 

Finally, let us consider the special case when the null hypersurface is an isolated horizon \cite{Ashtekar:2000sz,Ashtekar:2004cn,Engle:2010kt}: in this case the shear and expansion vanish, and the inaffinity is the surface gravity and it is constant. Then without choosing a foliation, the null symplectic potential reduces to
\begin{align}\label{thIH}
\Th^{\sscr IH}_{\scr g}(\d) = \int_{\cal N} 2 \om_{(n)\m} \d n^\m \, d{\cal N} +\int_{\p\cal N} \left(l^\m\d n_\m +l_\m \d n^\m\right) dS.
\end{align}
In NP notation, \Ref{defomega} reads
\be\label{omNP}
\om_{(n)\m} =  (\a+\bar\b)m_\m 
+ {\rm c.c.} +k_{(n)}l_\m,
\ee
and $(\a+\bar\b)$ is a normal-tangential component of $\na_\m n_\n$ on $\cal N$. This can be recognized as (minus) the rotational 1-form of the isolated horizons framework,\footnotemark
\be
\om_{(n)\m} = -\om_\m^{\sscr IH}.
\ee
\footnotetext{Recall that the pull-back on the connection index of the gradient of $n_\m$ gives
\be\nn
\underset{\leftarrow}{\na}{}_\m n_\n = \big[\g l_\m - (\a+\bar\b)m_\m \big]n_\n + (\l m_\m +\m \bar m_\m - \n l_\m) m_\n +{\rm c.c.}.
\ee
In this formula only, $\g$ is not our notation for the 2d metric determinant, but one of the NP spin coefficients, $k_{(n)}=-\g-\bar\g$. The term above proportional to $n_\n$ is the rotational 1-form, and coincides with minus \Ref{omNP}.
The term proportional to $m_\m$ vanishes under the isolated horizon conditions.
}

We see that our geometric decomposition \Ref{thEHn} correctly reproduces various special cases in the literature. As for the most general case, there will be also an additional contribution from $n^2\neq 0$ off $\cal N$ that is given at the end of the Section. 

Taking it into account, our decomposition should be compared with the one computed with metric variables in \cite{Parattu:2015gga}. 
We leave the question of their equivalence for future work.

Next, we show how this decomposition can be obtained in a quick and elegant way using tetrads.

\subsubsection*{Proof of Eq.~\eqref{thECnull2}}

Our starting point is the divergence of the normal, as in the non-null case. This can be written in tetrad variables as follows,
\begin{align}\label{tkforms}
\f12 \eps_{IJKL} ~e^I\w e^J\w  l^K d_\om n^L &\stackrel{\cal N}{=}  -\f12 \eps_{IJKL} ~e^I_\m e^J_\n  l^K D_\r n^L \eps^{\m\n\r\s}n_\s d{\cal N} \nn
\\ 
&= \na_\m n^\m d{\cal N} =  (\th_{(n)} + k_{(n)}) d{\cal N}.
\end{align}
Notice that the left-hand side is independent of the choice of $l$, namely invariant under \Ref{classII}.
Taking an arbitrary variation of the first expression we find 
\begin{align}\label{voeller}
\f12\eps_{IJKL}~&\d\big(\Si^{IJ} \w l^{K}d_{\omega}n^L \big) =\f12\eps_{IJKL}~\Sigma^{IJ}\w l^{K}\d \om^{LM}n_{M} + \f12\eps_{IJKL} \Big[ \d \Sigma^{IJ} \w l^K d_\om n^L 
 \nn\\
&\hspace{-8mm}+ \Sigma^{IJ}\w \left(\d l^{K}d_\om n^L + \d n^{K} d_{\om}l^L \right) -  d_\om \Si^{IJ} l^{K}\d n^{L}\Big] +\f12d\left(\eps_{IJKL}\Sigma^{IJ} l^{K} \d n^L  \right),
\end{align}
after an integration by parts to eliminate the derivatives on the variations.
As before, the first term in the right-hand side contains the symplectic potential, since
\begin{align}
 \f12 \eps_{IJKL}~\Si^{IJ}\w l^K\d\om^{LM} n_M &\stackrel{\cal N}{=} - n_I e^{\m}_{J}\d \om^{IJ}_{\m} d{\cal N} \stackrel{\cal N}{=} \f12\theta_{\scr e}.
\end{align}
For the second term, we use the inverse formula for the tetrad (see \Ref{eee}), finding
\begin{align}\label{uno}
\eps_{IJKL}~\d e^{I}\w e^J \w l^K d_\om n^L &\stackrel{\cal N}{=} - \big[(e^\r_I+l^\r n_I)\na_\r n^\m - (e^\m_I+l^\m n_I)\na_\r n^\r \big] \d e_\m^I d{\cal N}\nn\\ & = - \big[ \g^\r_I \na_\r n^\m-(\g^\m_I -n^\m l_I)\na_\r n^\r -k_{(n)}n^\m l_I \big] \d e_\m^I  d{\cal N},
\end{align}
where $\g^\m_I = \g^{\m\n} e_{I\n}$.
 The third and fourth terms combine together to give 
\begin{align}\label{due}
\f12&\eps_{IJKL}~ e^{I}\w e^J \w ( \d l^K d_\om n^L +\d n^K d_\om l^L)  \stackrel{\cal N}{=} \nn \\
& - \big[\d e_\m^I \big(\g^\n_{\;I}  n^\m l_\s \na_\n n^\s - k_{(l)} n^\m n_I  + k_{(n)}n^\m l_I  -n^\m l_I \na_\n n^\n\big)+ 2\d n^\m l_{[\n}\na_{\m]} n^\n \big]d{\cal N}.
\end{align}

We now restrict the variation to preserve the null nature of the hypersurface. This eliminates the term proportional to $k_{(l)}$ in \Ref{due},  since $\d e_\m^I n^\m n_I = -n_\m \d n^\m =0$.   
Adding up \Ref{uno} and \Ref{due}, we arrive at 
\begin{align}\label{pluto}
& \eps_{IJKL} \left[\d e^{I}\w e^J \w l^K d_\om n^L + \f12 e^{I}\w e^J \w ( \d l^K d_\om n^L +\d n^K d_\om l^L)\right] \nn \\
& \qquad \stackrel{\cal N}{=} - \d e_\m^I \Big[\g^\n_{\;I} \left(\na_\n n^\m + n^\m l_\s \na_\n n^\s\right) -\g^\m_{\;I} \na_\n n^\n\Big]d{\cal N} 
-  2 \d n^\m l_{[\n}\na_{\m]} n^\n d{\cal N}  \nn\\
& \qquad = \f12 \Big[B_{\m\n}\d \g^{\m\n}-(\th_{(n)}+k_{(n)})\g_{\m\n} \d \g^{\m\n} -2 \left(\g^\r_{\;\m} l^\s \na_\r n_\s-\th_{(n)} l_\m\right) \d n^\m
\Big]d{\cal N},
\end{align}
where we used the identities 
\begin{subequations}
\begin{align}
& B_{\m\n}\d\g^{\m\n} = \g^\r_\m \g^\s_\n \na_\r n_\s \d g^{\m\n} = -2 \g^\r_I \g^{\m\s} \na_{(\r} n_{\s)} \d e^I_\m = 
- 2 \g^\r_I (\na_{\r} n^{\m}+l^\s\na_\r n_\s n^\m) \d e^I_\m, \\
&\d\ln\sqrt{\sg}=-\f12 \g_{\m\n}\d\g^{\m\n}= \g^\m_I\d e_\m^I.
\end{align}
\end{subequations}
Finally for the boundary term, we have
\begin{align}
\f12\eps_{IJKL}~\Sigma^{IJ} l^{K} \d n^L  &\stackrel{\p \cal N}{=} \f12\eps_{IJKL}~e^{I}_\m e^{J}_\n l^{K} \d n^L \eps^{\m\n\r\s} l_\r n_\s dS
 = - l_I \d n^I dS.
\end{align}
Coming back to \Ref{voeller}, the variation of the final expression gives
\be \label{ertedesco1}
 \delta\Big((\th_{(n)} + k_{(n)}) d{\cal N} \Big)= \d (\theta_{(n)}+k_{(n)} ) d{\cal N} + (\theta_{(n)}+k_{(n)} ) \Big(\d \ln\f{\sqrt{-g}}f \Big) d{\cal N}.
\ee
Putting everything together, 
the last term in \Ref{ertedesco1} combines with the second term of \Ref{pluto} giving \Ref{ldn}, and we obtain
\begin{align}
\int_{\cal N} &\left[ \d (\theta_{(n)}+k_{(n)} ) \right]d{\cal N} = \f12 \Theta_{\scr e}(\d)  
+\f12\int_{\cal N} \big[B_{\m\n}\d \g^{\m\n} 
-2 \d n^\m (\g^\r_{\;\m} l^\s \na_\r n_\s +k_{(n) } l_\m)\big]d{\cal N}  \nn\\
&\quad \hspace{3.4cm} - \int_{\p\cal N} l_I \d n^I dS.
\end{align}
Solving for the symplectic potential, we get Eq.~\eqref{thECnull1}.

If $n^2=0$ only at $\cal N$, we get an additional $\tfrac12 l^\m\p_\m n^2 d\cal N$ in the right-hand side of \Ref{tkforms}, and an additional
$\tfrac12 (\d l^\m +e^\r_I\d e_\r^I l^\m)\p_\m n^2 d\cal N$ in the right-hand side of \Ref{pluto}. Assuming that $\d n^2=0$ and thus $n_\m\d n^\m=0$ still hold, the final result for the potential is then \Ref{thECnull1} with the additional term
\be
- \f12 \int_{\cal N}l^\m\p_\m n^2 \, (\d\ln f)\, d{\cal N}
\ee
on the right-hand side.

\section{Canonical pairs} \label{Sec:Pairs}

The symplectic potential can be used to read off the canonical pairs of the Lagrangian in a covariant way, without introducing an explicit Hamiltonian 3+1 decomposition. In this Section we look at the canonical pairs corresponding to the bare tetrad and metric-equivalent choices of symplectic potentials.
This will allow us to point out the differences in the corner pairs. The bulk pairs coincide,  but for null hypersurfaces it will nonetheless be useful to discuss them in some detail to understand the relation between various results in the literature.

\subsubsection*{Non-null hypersurfaces}

Introducing the unimodular induced metric $\hat q^{\m\n}:=(-sq)^{1/3}q^{\m\n}$, we can rewrite the bulk and boundary terms of the symplectic potentials \eqref{maggica1}-\eqref{thEH1} as
\begin{align}
\Th^\Si_{\scr e}(\d) &= \Th^\Si_{\scr g}(\d) \nn\\\label{Spairs}&
= s\int_\Si \left[\f43 K\d\ln\sqrt{-sq} + (-sq)^{-1/3} \left(K_{\m\n}-\f13 K q_{\m\n}\right)\d \hat q^{\m\n}\right]d\Si -2\d (Kd\Si) \\
\Th^{\p\Si}_{\scr e}(\d) &=  -2 \int_{\p\Si} \hat r_I  \d n^I dS,  \\
\Th^{\p\Si}_{\scr g}(\d) &=  - \int_{\p\Si} \hat r_\m  \d n^\m dS. 
\end{align}
In the bulk of the Cauchy hypersurface, we have two canonical pairs, as in the Hamiltonian formalism: the traceless part of the extrinsic curvature conjugated to the conformal metric, and the trace of the extrinsic curvature conjugated to the conformal factor. 

The symplectic potential  provides us also with a canonical pair on the boundary of the Cauchy hypersurface.
This pair is affected by the cohomology ambiguity, and the result is different with tetrad or metric variables. 
In both case, the conjugated pair is given by the normal to the hypersurface $\Si$ and the densitized (by the area density $\sqrt \g$) normal of the boundary within $T\Si$. 
The differences are a factor of 2 and the nature of the vectors: internal vectors with the bare tetrad potential versus spacetime vectors with the metric potential. In other words, it includes a non-trivial variation of (a projection of) the tetrad.

\subsubsection*{Null hypersurfaces}

On null hypersurfaces it is possible to give a finer identification of the canonical pairs, if one introduces a foliation along the null generators, say by the level sets of $r$ as before, and adapts the NP tetrad. We can then introduce a uni-modular induced metric on the 2d slices, 
\begin{align}
& \hat\g^{\m\n}:=\sqrt{\g} \g^{\m\n}, \qquad \check\g_{\m\n}=\g^{-1/2}\g_{\m\n}.
\end{align}
This allows us to diagonalize the shear and expansion contained in $B_{\m\n}$,
\begin{align}
\label{metricshear}
B_{\m\n}\d\g^{\m\n} = \s_{(n)\m\n}\d\hat\g^{\m\n} + \th_{(n)}\d\ln\sqrt{\g}, \qquad \s_{(n)\m\n}=\f12\pounds_n \check\g_{\m\n}. 
\end{align} 
Using this decomposition in the symplectic potentials on a null hypersurface, 
we can rewrite the bulk and boundary terms \eqref{thECnull1} and \eqref{thEHn} as
\begin{align}
\Th^{\cal N}_{\scr e}(\d) &=\Th_{\scr g}(\d) \nn\\ \label{thEHn2} &= \int_{\cal N} \big[  -\s_{(n)\m\n} \d \hat\g^{\m\n} 
+ \d ( \th_{(n)} + 2 k_{(n)}) +2 \om_{(n)\m} \d n^\m \big] d{\cal N}+ \int_{\cal N}\d(\th_{(n)}d{\cal N}), \\
\Th^{\p{\cal N}}_{\scr e}(\d) &= 2\int_{\p\cal N} l_I\d n^I  dS, \\
\Th^{\p{\cal N}}_{\scr g}(\d) &= \int_{\p\cal N} (l^\m\d n_\m + l_\m\d n^\m) dS. \label{nullBTg}
\end{align}

Let us first discuss the bulk part.
The first term is the well-known spin-2 pair made by the shear and the conformal 2d metric.\footnote{The fact that the momentum of the spin-2 configuration variables is a spatial derivative and not a velocity is the gravitational equivalent of the light-cone constraint of light-front field theory in Minkowski spacetime. In the first-order formalism, this crucial relation appears as a second class secondary constraint \cite{IoSergeyNull}.}
 It dates back to Sachs' initial value problem \cite{Sachs62}, 
shows up in the canonical analysis \cite{Torre:1985rw,Reisenberger:2007ku,Reisenberger:2018xkn,IoElena}, and features prominently in Ashtekar's symplectic structure at $\scri$ \cite{Ashtekar:1981bq} -- where all other terms vanish.
The second term is the spin-0 pair, made by the 2d conformal factor and its momentum. The latter is given by the combination $\th_{(n)}+2k_{(n)}$, and appears in this form also in the canonical formalism \cite{Torre:1985rw,IoElena} and in the analysis of \cite{Hopfmuller:2016scf}. 
Notice the factor of 2, which introduces a mismatch between the spin-0 momentum and the divergence of the hypersurface normal, which was the result in the space-like case. 
The third term has three components because of \Ref{ndn}, and describes a spin-1 pair, given by the tangent to the null generators and the rotational 1-form of the isolated horizon framework.  From the canonical perspective of \cite{Torre:1985rw}, the two components in $\eta_\m$ describe the momentum to one of the shift vectors of the 2+2 formalism, whereas the inaffinity piece would be the canonical momentum to the gauge-fixed metric component $g^{vv}$.\footnote{And requires the extension of the phase space mentioned in a previous footnote, see also discussion in \cite{IoSergeyNull}.}

Having fixed a foliation, we can now choose the canonical normalization \Ref{f=N}, which results in $d{\cal N}=\sqrt{\g}drd^2\th$ and $l_\m\d n^\m=0$, eliminating one component of the spin-1 momentum. The bulk symplectic potential is now identical to the one derived in  \cite{Torre:1985rw,Reisenberger:2007ku,Hopfmuller:2016scf}. 
If one further fixes the partial Bondi gauge, $\d n^\m=0$ and the spin-1 piece drops out of the potential entirely. It remains the gauge freedom to choose the $r$ coordinate, freedom which affects for instance the inaffinity and its factor of 2 in the spin-0 momentum. Choosing $r$ to be an affine parameter, the inaffinity variation vanishes, and $\th_{(n)}\equiv \p_r \sqrt\g$: the spin-0 term becomes entirely a corner term, indeed, one of Sachs' corner data \cite{Reisenberger:2007ku}.
We refer the reader to \cite{Hopfmuller:2016scf} for further discussions of the relevance of the spin-1 and spin-0 pairs.

In this discussion we used a foliation of $\cal N$ to separate the spin-0 and spin-2 parts of the canonical pairs, but in the special case of a non-expanding horizon, $\th_{(n)}=0$ and one can identify the pairs without introducing a foliation. This is for instance the case of future null infinity where only the spin-2 shear pair appears \cite{Ashtekar:1981bq,Ashtekar:2018lor}, or of an isolated horizon where only the spin-1 pair appears \cite{Ashtekar:2000sz,Ashtekar:2004cn}, see \Ref{thIH}.\footnote{ {While the formalism at future null infinity does not depend on a choice of foliation, it
does depend on a choice of normalization for the tangent vector, or in
other words, on a choice of conformal factor in the compactification.
For recent work aiming at a purely
conformal invariant description, see \cite{Herfray:2020rvq}.}}

An interesting remark about the canonical pairs on a null hypersurface is that {all momenta have a connection interpretation} in the first-order formalism:
In particular the shear and spin-0 momenta are the components of the connection on the little group of a null direction, respectively the null traslations for the shear, and the helicity rotation for the spin-0 momentum  $\th_{(n)}+2k_{(n)}$ \cite{IoElena}. This is in contrast with the case of non-null hypersurfaces, the momenta are component of the extrinsic curvature, which in the first-order formalism are hypersurface-orthogonal parts of the Lorentz connection, and do not transform as a connection under the little group preserving the hypersurface normal. This is the reason why one needs either to complexify the variables or to introduce (the term in the action proportional to) the Barbero-Immirzi parameter in order to achieve a connection formulation.
The connection interpretation of the momenta is a remarkable characteristic of a null hypersurface, and for the shear part it dates back to the seminal work at future null infinity \cite{Ashtekar:1981hw,Ashtekar:2018lor}.

\section{Variational principle and corner terms in the action} \label{Sec:Corners}

An advantage of using tetrads is that it is much simpler to join the boundary terms and obtain the variational principle on a closed region of spacetime with (non-orthogonal) corners, as pointed out in \cite{Jubb:2016qzt}. There the authors restricted attention to the variations preserving the induced metric on the boundary, and adapted the tetrad. Here we consider arbitrary variations to be able to deal with any boundary conditions one may choose for the variational principle.
In the context of joining boundaries, we can ignore the dressing 2-form $d\a(\d)$ since it is globally exact and therefore its contributions cancel out. The boundary terms to be added to the action for the variational problem are thus equivalent in metric or tetrad variables, the advantgae of using the latter is only that it simplifies evaluating them.

\subsection{Joining the boundaries}

To keep the presentation brief, we present only the two most common cases of finite regions, namely a time-like cylinder and a section of a light-cone.
Formulas for the other types of joints considered in the literature, \emph{e.g.} in \cite{Lehner:2016vdi, SorkinCorner17}, can be easily derived as below, paying the necessary attention to signs and relative orientations.
\begin{figure}[h!]
\def\svgwidth{\textwidth}
\centering 
\scalebox{0.4}{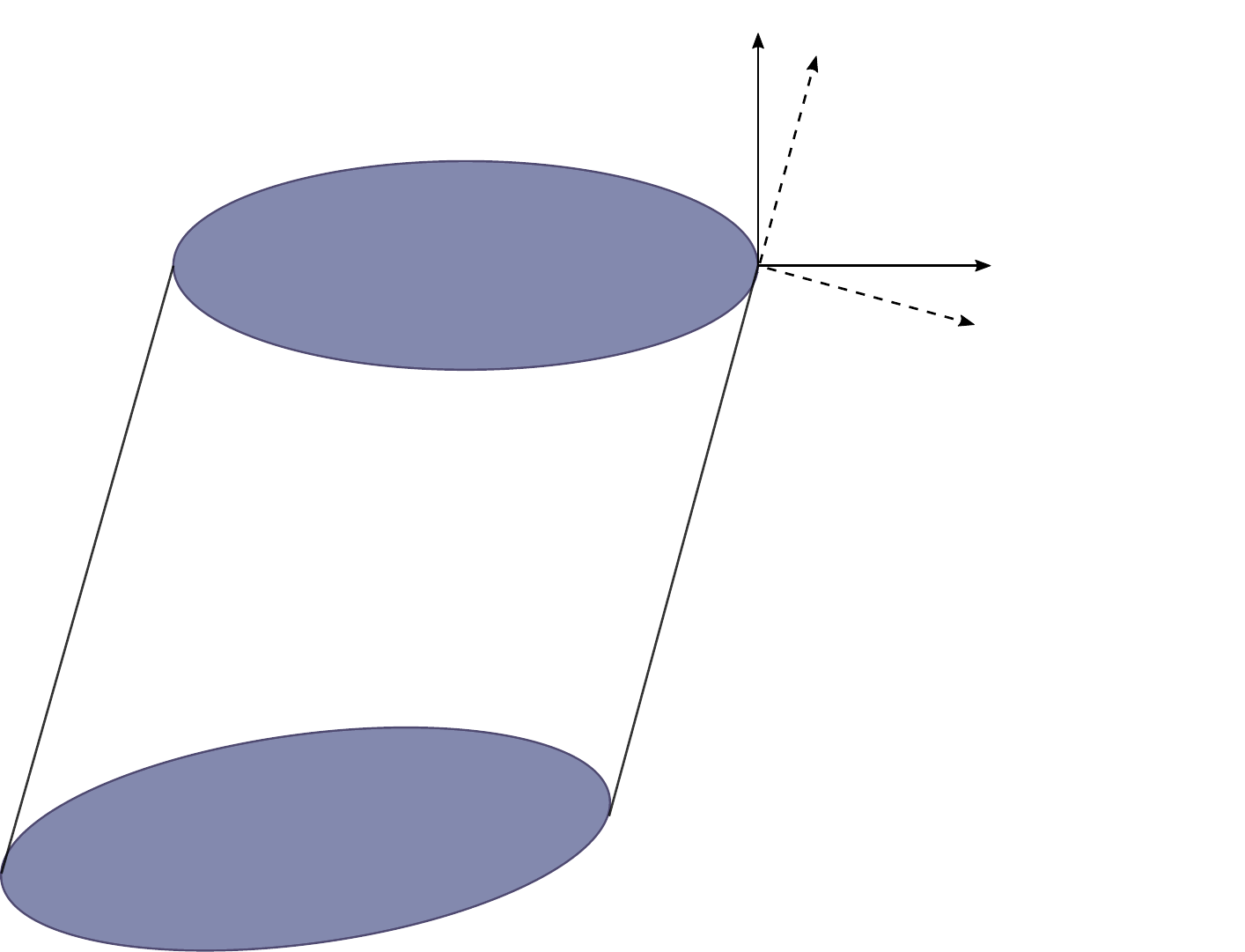} \hspace{0.5cm}
\scalebox{0.4}{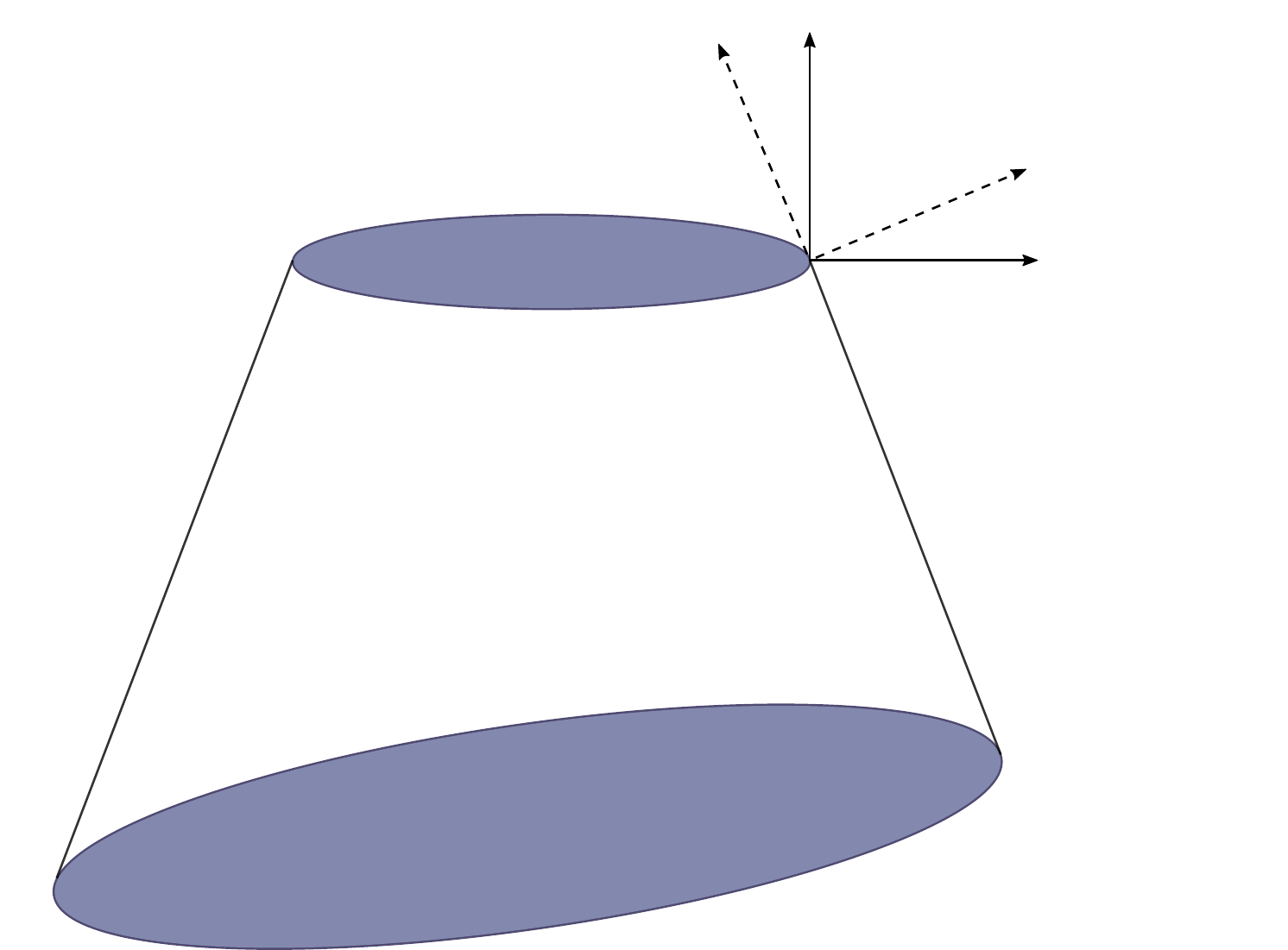}
\caption{\label{Fig}
\small{\emph{The two standard settings for the variational problem considered in this section.} \\
Left panel: \emph{The boundary of the four-dimensional domain of integration consists of a pair of space-like hypersurfaces $\Si_{1,2}$ and a time-like one $\cal T$, joined at the space-like corners ${\cal C}_{1,2}$. The figure shows the two basis $(n,\hat r)$ and $(r,\hat n)$, where $n$ is the (time-like) unit-norm normal to $\Si_2$, $r$ is the (space-like) unit-norm normal to $\cal T$, while $\hat n$ and $\hat r$ are, respectively, the unit-norm projections of $n$ and $r$ in T$\cal T$ and T$\Si$.}  \\
Right panel: \emph{The boundary of the four-dimensional domain of integration consists of a pair of space-like hypersurfaces $\Si_{1,2}$ and a converging section of a past light-cone $\cal N$, joined at the space-like corners ${\cal C}_{1,2}$. 
The figure shows the two basis $(\tau,\hat r)$ and $(n,l)$, where $\tau$ is the (time-like) unit-norm normal to $\Si_2$, $n$ is the (null) normal to $\cal N$, while $\hat r$ and $l$ are, respectively, the unit-norm projections of $n$ in T$\Si$ and the transverse null vector to $n$.} 
} }
\end{figure}
We consider first the time-like cylinder, left panel of Fig~\ref{Fig}. The outgoing time-like normal is future-pointing at  $\Si_2$, and past-pointing at  $\Si_1$. This gives a global minus sign when applying the formulas of Section~\ref{SecNN} to $\Si_1$. As for the time-like boundary $\cal T$, the outgoing unit normal is space-like and we denote it by $r$. We use $\hat r$ for the unit-norm projection of $r$ in $T\Si$, and $\hat n$ for the unit-norm projection of $n$ in $T{\cal T}$. Using \Ref{maggica1}, 
the total on-shell variation of the tetrad action is
\begin{align}\label{totald1}
\d S_{\scr e} &\approx \int_{\Si_2}\th(\d)-\int_{\Si_1}\th(\d) +\int_{\cal T}\th(\d) \\\nn
&=\int^{\Si_2}_{\Si_1}(2\d K-K_{\m\n}\d q^{\m\n})d\Si - \int_{\cal T}(2\d K-K_{\m\n}\d q^{\m\n})d\Si  - 2 \int^{C_2}_{C_1} (\hat r_I \d n^I +\hat n_I\d r^I) dS. 
\end{align}
To evaluate the corner term, we follow the standard procedure (see, \emph{e.g.}, \cite{Sorkin:1975ah,Hartle:1981cf,Hayward:1993my,Hawking:1996ww}, and more recently \cite{Lehner:2016vdi,Jubb:2016qzt}) and introduce an $\SO(1,1)$ transformation between the basis $(n,\hat r)$ and $(r,\hat n)$,
\be
n_I = \sinh \eta \,r_I + \cosh \eta \,\hat n_I, \qquad \hat r_I = \cosh\eta \,r_I +\sinh\eta \,\hat n_I.
\ee
Here, $\sinh\eta=n_I r^I$ is the boost between the two hypersurfaces, and it vanishes in the case of orthogonal corners. 
Using these formulas and their inverses, we have
\be\label{deta}
\hat r_I \d n^I +\hat n_I\d r^I = \f1{\cosh\eta}(r_I \d n^I + n_I\d r^I) = \d\eta.
\ee
Plugging this in \Ref{totald1}, we find the corner contribution $-2\d\eta \, dS$. This can be  recognized as the non-orthogonal corner contribution to the variation of the EH action \cite{Hayward:1993my}. It can also be derived from the metric boundary terms of \Ref{thEH1}: 
\be \label{deta2}
\hat r_\m \d n^\m +\hat n_\m\d n^\m = 2(\hat r_I \d n^I +\hat n_I\d n^I) = 2\d\eta.
\ee
The one-line derivation  of the corner term can be advantageously compared with the rather longer one required in metric variables; see, \emph{e.g.}, \cite{Hayward:1993my,Lehner:2016vdi}.

In our second example, we replace the time-like boundary with a section of a past light-cone; see right panel of Fig.~\ref{Fig}.
To avoid a notational problem, we now use $\t^\m$ for the future-pointing time-like normal, and keep $n^\m$ for the null normal to $\cal N$. In the adapted NP tetrad $(l,n,m,\bar m)$, the integrable vectors $(m^\m,\bar m^\m)$ foliate $\cal N$, and $l^\m$ is tangent to light rays outgoing from $\cal N$. Using \Ref{maggica1} and \Ref{thECnull1} 
the total on-shell variation of the tetrad action is
\begin{align}\label{totald2}
\d S_{\scr e} &\approx \int_{\Si_2}\th(\d)-\int_{\Si_1}\th(\d) +\int_{\cal N}\th(\d)\nn \\ \nn
&=\int^{\Si_2}_{\Si_1}(2\d K-K_{\m\n}\d q^{\m\n})d\Si +
\int_{\cal N} \left[ -B_{\m\n} \d \g^{\m\n} 
+ 2 \d \left( \th_{(n)} + k_{(n)}\right) +\om_{(n)\m} \d n^\m \right] d{\cal N} 
\\&\qquad 
+ 2 \int^{C_2}_{C_1} (-\hat r_I \d \t^I + l_I\d n^I) dS. 
\end{align}
The change of basis at the corners is given by \cite{Jubb:2016qzt}
\be
\t^I = \f1{\sqrt 2}(e^\eta l^I+ e^{-\eta} n^I), \qquad \hat r^I = \f1{\sqrt 2}(e^\eta l^I-  e^{-\eta} n^I),
\ee
where $\t_I n^I=-e^\eta/\sqrt 2 $ measures the boost among the normals, and it vanishes when $\t^I=(l^I+n^I)/\sqrt2$.
From these relations it follows that
\be
-2 \hat r_I\d\t^I +2l_I\d n^I =-2\d\eta,
\ee
and thus the corner contribution to the variation is $-2\d \eta\, dS$.
In terms of spacetime tensors, using \Ref{silva} and \Ref{statecollege}, one finds that
\be
-2 \hat r_I\d\t^I +2l_I\d n^I = -\hat r_\m\d \t^\n + l^\m \d n_\m + l_\m\d n^\m.
\ee
Comparing the last two equations, we have recover the corner term derived in \cite{Lehner:2016vdi,Jubb:2016qzt}, here generalized without any coordinate or internal Lorentz gauge fixing, and the associated restricted variations.
Again, we would like to stress the simplicity and the generality of the derivation performed with tetrads variables.

\subsection{Variational problems with non-orthogonal corners}

The tetrad and metric Lagrangians \Ref{SEC} and \Ref{LEH} give a well-defined variational principle with boundary conditions fixing the extrinsic curvature, or the projected connection in the first-order formulation. Dirichlet boundary conditions fix instead the induced metric, and in this case one has to supplement the Lagrangians with boundary terms to have a well-defined variational principle. The required boundary terms can be read off the formulas of the previous Section imposing the Dirichlet condition $\d q_{\m\n}=0$ at the boundary. 

If the boundary is the time-like cylinder of Fig.~\ref{Fig}, left panel, we have 
\begin{align}\label{D1}
 \d^{\sscr D} S_{\scr e} &\approx \int^{\Si_2}_{\Si_1} 2\d K d\Si - \int_{\cal T}2\d K d\Si - \int^{C_2}_{C_1} 2\d\eta \, dS
\nn\\& = 2\d \left(\int^{\Si_2}_{\Si_1} K d\Si - \int_{\cal T} K d\Si - \int^{C_2}_{C_1} \eta dS\right) =: - \d I_{3d} - \d I_{2d}.
\end{align}
This is a total variation, and we can identify the usual Gibbons-Hawking and Hayward boundary terms. 
In tetrad variables, these can be written as
\begin{subequations}
\begin{align}
I_{3d}^{\sscr D} &= s ~2K d\Si = s~2\na_\m n^\m d\Si = s~\eps_{IJKL}~e^I\w e^J \w n^K d_\om n^L, \\
I_{2d}^{\sscr D} &= 2\eta dS = 2\sinh^{-1} (n_I r^I) dS  = \f12\sinh^{-1} (n_I r^I) \, \eps_{IJKL}~e^{I}\w e^J n^K r^L.
\end{align}
\end{subequations}
Recently, it has also been suggested to consider mixed boundary conditions $\d K=0=\d\hat q_{\m\n}$  \cite{Anderson:2010ph,Witten:2018lgb,Wieland:2018ymr}, where $\hat q_{\m\n}$ is the unimodular representative of the conformal class of induced metrics. The boundary variation in this case can be read from \Ref{Spairs}, and the only difference is a numerical factor in the 3d boundary term to be added to the action,
\be
I_{3d}^{\sscr mixed} = s \f 23\int_{\Si} K d\Si.
\ee

For the past light-cone of Fig.~\ref{Fig}, right panel, Dirichlet boundary conditions impose $\d \g_{\m\n}=0$ on the null boundaries. Notice that, when the connection is on-shell in the absence of torsion, this also implies that $\d \th_{(n)}=0$.\footnote{See, \emph{e.g.}, \cite{Speziale:2018cvy} for the contribution of the torsion to the expansion, and to null geodesic congruences in general.} Then,
\begin{align}
 \d^{\sscr D} S_{\scr EC} &\approx \int^{\Si_2}_{\Si_1} 2\d K d\Si - \int_{\cal N}(2\d k_{(n)} +\om_{(n)\m} \d n^\m)d{\cal N} -  \int^{C_2}_{C_1} 2\d\eta dS,
\end{align}
but this is not a  total variation. The obstruction is the $\d n^\m$ term. Since $n^\m$ is tangent to the null hypersurface, its variation  must be also fixed in order to have a well-defined variational principle, as discussed in \cite{Parattu:2015gga,Lehner:2016vdi,Jubb:2016qzt}.
This can be achieved working in partial Bondi gauge, which fixes $\d n^\m=0$ while still allowing variations of the inaffinity. In this gauge,
\begin{align}
 \d^{\sscr D+Bondi} S_{\scr EC} &\approx \int^{\Si_2}_{\Si_1} 2\d K d\Si - \int_{\cal N}2\d k_{(n)} d{\cal N} -  \int^{C_2}_{C_1} 2\d\eta dS
\\\nn& = 2\d \left(\int^{\Si_2}_{\Si_1} K d\Si - \int_{\cal N} k_{(n)} d{\cal N}- \int^{C_2}_{C_1} \eta dS\right) =: - \d I_{3d} - \d I_{2d}.
\end{align}
This is now a total variation, and we can identify the boundary terms to add to the action. The space-like ones are the same as before. The 3d null boundary term is $2k_{(n)}d{\cal N}$, but since $\d\th_{(n)}=0$ with Dirichlet boundary conditions, it can also be taken to be the following covariant tetrad expression,
\begin{align}
I^{\sscr D+Bondi}_{3d} &=2(\th_{(n)}+k_{(n)})d{\cal N} = 2\na_\m n^\m d{\cal N} =  \eps_{IJKL}~e^I\w e^J \w l^K d_\om n^L.
\end{align}
As for the new corner term, in tetrad variables this reads
\begin{align}
I^{\sscr D}_{2d} &=2\eta dS = 2\ln(-\sqrt 2\, n_I \t^I) dS  = \f12 \ln(-\sqrt 2\, n_I \t^I)  \, \eps_{IJKL}~e^{I} \w e^J \, l^K n^L.
\end{align}

We remark that the 3d boundary term has a universal form for all hypersurfaces when written as the divergence of the hypersurface normal.
The expression in the null case is however coordinate-dependent, because the inaffinity depends on the choice of the $r$ coordinate. Only upon completing the gauge-fixing, one has an unambiguous expression. The simplest choice is $r$ an affine parameter, for which the boundary term vanishes.
Alternatively, this dependence of the boundary action on the parameters of the null generators could be eliminated adding an additional corner term \cite{Lehner:2016vdi,Jubb:2016qzt}, but it was also argued not to affect physical quantities \cite{Jubb:2016qzt}.

\section{Hamiltonian surface charges} \label{Sec:charges}

Theories with local gauge symmetries, like Yang-Mills and General Relativity, admit an elegant extension of Noether's theorem, that shows how the charges -- relevant for conservation or balance laws -- are associated with surface integrals. 
These surface charges date back to the ADM analysis \cite{ADM} and subsequent literature; see \emph{e.g.}, \cite{Regge:1974zd,AshtekarHansen,Beig:1994ya}.
A convenient modern framework to address this question is the covariant phase space prescription \cite{WittenCovPhase,AshtekarReula,LeeWald}, whose mathematical structure was later also associated with a variational bi-complex or jet bundle \cite{Anderson:1996sc,Barnich:2000zw}.
The literature slightly branches off three different viewpoints, depending on the mathematical tools one focuses on: the (pre)-symplectic 2-form, the Noether charge from the symplectic potential, or  Anderson's homotopy operator. This branching, together with the use of similar names for different quantities, can create some confusion in comparing results. 
For this reason we decided to include a brief introduction to fix notations and conventions, even though the material is well-covered in the literature.
As a slight generalization to much of the literature, we allow throughout for field-dependent gauge parameters. This has relevance for questions of integrability or multipole decompositions.
We hope not to alienate the reader familiar with these notions, and  invite her to skip the next subsection.

\subsection*{Covariant phase space methods}\label{CPS}

In covariant phase space methods \cite{Ashtekar:1981bq,WittenCovPhase,LeeWald,AshtekarReula,Iyer:1994ys,Wald:1999wa,Ashtekar:2008jw}, one associates a symplectic potential to a hypersurface $\Si$ from the integral \Ref{Thdef} of the boundary variation of the Lagrangian.
We follow \cite{Wald:1999wa,Ashtekar:2008jw} and denote with $\d$ a specific functional variation, corresponding to a vector field over the field space $\cal F$; and with $\Th(\d)= I_\d \Th$ is the inner product in $\cal F$ between a 1-form and a vector field. The pre-symplectic 2-form is the exterior derivative in field space, $\Om:=\fd \Th$. It can be written in terms of standard functional differentials as follows,
\be\label{defOm}
\Om(\d_1,\d_2) = \d_1[\Th(\d_2)] - \d_2[\Th(\d_1)] -\Th([\d_1,\d_2]).
\ee
This quantity depends a priori on the hypersurface $\Si$ chosen to evaluate the integrals, but it can be easily shown to be closed in spacetime, $d\Om\approx 0$,
if the fields and their linear variations are on-shell; see, \emph{e.g.}, \cite{LeeWald}. 

Having a pre-symplectic form at disposal, one can look for the Hamiltonian generator associated with a symmetry $\d_\eps$
\be\label{defHeps}
\sd H_\eps := \Om(\d,\d_\eps) = \d[\Th(\d_\eps)] - \d_\eps[\Th(\d)] - \Th([\d,\d_\eps]).
\ee
The slashed delta used in this definition is meant to highlight that the right-hand side is not necessarily a total functional variation.
When it is, we say that the expression is integrable, and refer to $H_\eps$ as the Hamiltonian generator. 
Simple sufficient conditions for integrability are
\be
\d_\eps[\Th(\d)] =0, \qquad  [\d,\d_\eps]=0.
\ee
It can be shown by explicit calculation that for internal gauge and diffeomorphism symmetries, the integrand in \Ref{defHeps} is exact. Therefore $\sd H_\eps$ is a surface integral, if $\Si$ has a single boundary, or the difference of two surface integrals  if $\Si$ has two disconnected boundaries. When the generator is integrable, we will then refer to $H_\eps$ on one boundary as the Hamiltonian charge.

These charges and their fluxes are the main object of interest in covariant phase space methods.
A proof that on general grounds these quantities are always surface integrals was given in \cite{Iyer:1994ys}.  Because of this proof, some literature refers to the Hamiltonian charge also as Iyer-Wald charge.
We briefly sketch the proof, which will allow us to introduce a second quantity of interest, the Noether charge, and to better appreciate the issue of integrability.
We consider arbitrary gauge parameters, including field-dependent  ones and, accordingly, we keep track of terms in $\d\eps$ and $[\d,\d_\eps]$.

The starting point is the variation of the Lagrangian under a gauge symmetry. By definition of Lagrangian symmetry, this can be at most a boundary term,
\be\label{depsL}
\d_\eps L = \FE{\phi}\d_\eps\phi +d\th(\d_\eps) = d Y_\eps.
\ee 
The quantity
\be\label{jW}
j(\d_\eps):=\th(\d_\eps) - Y_\eps
\ee
is closed on-shell, \emph{i.e.}, $dj(\d_\eps) \approx 0$, and can be taken as definition of Noether current for the symmetry $\eps$ \cite{Iyer:1994ys}.
For gauge symmetries, this current is also exact on-shell (see, \cite{Iyer:1994ys,Wald:1999wa} for the general proof and below for explicit cases), and we denote
\be
j(\d_\eps) \approx d q_\eps.
\ee
The associated surface integral 
\be\label{NoetherIW}
Q_\eps:= \int_\Si j(\d_\eps) \approx \int_{\p\Si} q_{\eps}
\ee
is referred to as Noether charge \cite{Iyer:1994ys}; we will follow this naming.
A straightforward calculation then gives 
\be\label{HepsG}
\sd H_\eps \approx \int_\Si \left(d \d q_\eps - \d_\eps\th(\d) - dq_{\d\eps}+\d Y_\eps - Y_{\d\eps}\right).
\ee

We now distinguish the two cases of internal gauge symmetries and diffeomorphisms. For internal gauge symmetries, which we denote with $\eps=\l$, the variation of the Lagrangian is exactly zero, so $Y_\l=0$. The symplectic potential is gauge-invariant, meaning $\d_\l\th(\d)=0$.
We have that the Noether current coincides with the symplectic potential, 
 and \Ref{HepsG} reduces to 
\be\label{Hl}
\sd H_\l \approx \int_{\p\Si} \left(\d q_\l -q_{\d\l}\right).
\ee
For field-independent gauge parameters, the Hamiltonian generator associated to internal gauge symmetries is integrable and it coincides with the Noether charge \eqref{NoetherIW}.

For diffeomorphism, which we denote with $\eps=\xi$, the variation of a generally covariant Lagrangian
gives $Y_\xi=i_\xi L$. 
The Noether current does not coincide with the symplectic potential, and \Ref{HepsG} gives
\be\label{Hxi}
\sd H_\xi \approx \int_{\p\Si} \left(\d q_\xi - i_\xi\th(\d) - q_{\d\xi}\right).
\ee
The integrability of this expression requires a case by case study.

It is useful to make these expressions more concrete with two notable examples: the Yang-Mills and Einstein-Hilbert Lagrangians. 
The first is given by $L=-\f12 \tr (F\w\star F)$, with symplectic potential
$
\th(\d)=-\tr (\d A\w \star F). 
$
The Lagrangian is invariant under a gauge transformation $\d_\l A:=-d_A\l$, and
\be\label{jYM}
j(\d_\l):=\th(\d_\l) = d \, \tr \left(\l \star F\right). 
\ee
Hence, the Noether charge is
\be\label{QYM}
Q_\l := \int_{\Si} j(\d_\l) = \int_{\p\Si}\tr \left(\l\star F\right).
\ee
We further observe from \Ref{jYM} that for a covariantly constant gauge transformation, $d_A\l=0$, the Noether current vanishes on-shell in vacuum: in this case the Noether charge obeys a 3d conservation law, namely its value  
is independent of deformations of the integration surface $\p\Si$. This is the standard case of interest for charges.
To compute the Hamiltonian generator, we first observe that the symplectic potential is gauge-invariant, $\d_\l\th(\d)=0$. We then have
\be\label{Hym}
\sd H_\l =\d[\Th(\d_\l)] - \Th([\d,\d_\l]) \approx \int_{\p\Si}\tr \left(\l\star\d F \right).
\ee
For field-independent gauge parameters, $\d\l=0$ and $[\d,\d_\l]=0$. The Hamiltonian generator is then integrable, and the Hamiltonian charge coincides with the Noether charge \Ref{QYM}. 

For the second example, the Einstein-Hilbert Lagrangian \Ref{LEH} has the symplectic potential \Ref{thg}, and 
\begin{align}\label{jEH}
j(\d_\xi):= \th(\d_\xi)-i_\xi L &= 
d\k_\xi + \FE{e}_I\xi^I \approx d\k_{\xi},
\end{align}
where $(\FE{e}_{I})_{\m\n\r} = 2 \eps_{\m\n\r\a}(G^{\a\b} +\L g^{\a\b})e_{\b I}\approx 0$, and
\be\label{Komar}
\k_\xi:=-\f12 \eps_{\m\n\r\s} \na^\r\xi^\s dx^\m\w dx^\n
\ee
is the Komar $2$-form. The {Noether charge} for diffeomorphisms is the surface integral of the Komar 2-form,
\be\label{Qdiff}
Q_\xi:=\int_\Si j(\d_\xi) \approx \int_{\p\Si}\k_\xi.
\ee
As for the YM case, this expression becomes independent of the integration surface $\p\Si$ only in vacuum and in the case of isometries, namely when $\xi$ satisfies the Killing equation. This is proved using 
$d\k_\xi= \star (2\na^\n \na_{[\n}\xi_{\m]}dx^\m) = -2 \star (R_{\m\n}\xi^{\n}dx^\m)$ for a Killing vector. Hence, the Noether current \eqref{jEH} vanishes in vacuum and for an isometry.\footnote{Excluding the presence of singularities.} 

The explicit form of the Hamiltonian generator \Ref{Hxi} is 
\begin{align}
\sd H_\xi &\approx \int_{\p\Si} \left(\d\k_\xi -i_\xi \th(\d) - \k_{\d\xi}\right) \nn\\
&= - \f12 \int_{\p\Si} \eps_{\m\n\r\s}\big[ (\d\ln{\sqrt{-g}}) \na^\r\xi^\s + \d g^{\r\a}\na_\a\xi^\s 
+ \xi^\r \left(\na_\a \d g^{\a\s} + 2 \na^\s \d\ln\sqrt{-g}\right) \nn\\ &\hspace{3cm} - \xi_\a\na^\r\d g^{\s\a} \big] dx^\m\w dx^\n. \label{IWmetric}
\end{align}
The integrability of this expression is non-trivial. The only simple case concerns field-independent diffeomorphisms tangential to $\p\Si$, for which the second and third terms in the first equality vanish. The generator is then manifestly integrable, and the Hamiltonian charge coincides with the Noether charge given by the Komar 2-form.
For non-tangential diffeomorphisms the situation is more subtle and one has to do a case by case study; see \emph{e.g.}, time diffeomorphisms and the ADM energy at spatial infinity, which is not given by the Komar 2-form alone.
The situation should be compared with the Noether charge \Ref{Qdiff}, which is always well-defined. However, it is  the Hamiltonian charge that generates the symmetry in phase space. Furthermore, as we recall next the definition of the Noether charge used above is ambiguous, since it is changed adding boundary terms to the Lagrangian, whereas the Hamiltonian charge it is not.

As before, the most relevant situation for physical applications concerns the case of isometries (be them global or asymptotic), for which the value of the charges doesn't depend on the surface of integration chosen in vacuum. Nonetheless, we will keep a general mind in this paper and consider all charges, including those not associated with isometries, and thus a priori surface-dependent.

\subsubsection*{Ambiguities}

There are two ambiguities in the definition of the symplectic potential \cite{Iyer:1994ys}:
\begin{itemize}
\item[$I$.] \emph{Boundary terms}: if one adds a boundary term to the Lagrangian (without introducing new fields), \emph{i.e.}, $L\mapsto L+dY$, then $\th\mapsto\th+\d Y$. This is like a change of polarization in the phase space (\emph{e.g.}, $pdq$ going to $-qdp$). It affects the symplectic potential and the value of the Noether charge, but does not affect the symplectic structure and Hamiltonian charges.
\item[$II$.] \emph{Cohomology ambiguity}: the Lagrangian gives a unique prescription for $d\th$. In extracting the symplectic potential $\th$, one is always free to add an exact form to it, so the symplectic potential is only defined up to the cohomology ambiguity $\th\mapsto \th+d\a$ for an arbitrary 2-form $\a$.  This ambiguity \emph{does} affect  the symplectic structure and Hamiltonian charges, as well as the symplectic potential and Noether charges. 
\end{itemize}
A third ambiguity of covariant phase space methods is the fact that the surface charge itself is only defined up to the addition of an exact 2-form, but this is often irrelevant since attention is restricted to compact (but not necessarily connected) $\p\Si$.

The ambiguity $II$ plays an important role in our considerations about tetrad GR. To highlight it, we call \emph{bare} the symplectic potential that  can be read off the variation of the Lagrangian without any additional inputs. As discussed in  \cite{DePaoli:2018erh} and more extensively below, the Noether and Hamiltonian charges obtained from the bare tetrad symplectic potential differ from those obtained in metric variables, even though the space of physical solutions is the same. This is a direct consequence of the mismatch \Ref{thda}. It can then be compensated exploiting the ambiguity $II$ to dress the bare symplectic potential with the dressing 2-form \Ref{DPS}, thus restoring equivalence with the metric charges.

Another situation in which the ambiguity $II$ has been used to define an improved symplectic potential is in the context of renormalization of the super-Lorentz charges at null infinity  \cite{Compere:2018ylh}. Dressing the symplectic potential with an exact form can be also obtained  adding a boundary term that depends on new fields  to the initial Lagrangian, as done in  \cite{DonnellyFreidel,Wieland:2017zkf,Geiller:2017whh,Wieland:2019hkz}. In particular the results of \cite{Wieland:2017zkf,Wieland:2019hkz} indicate that the main properties of our dressed tetrad symplectic potential, namely zero Lorentz charges and metric-equivalent diffeomorphism charges, can be obtained with a boundary Lagrangian describing the induced geometry on a null boundary in terms of spinors.

\subsection{Bare and gauge-invariant pre-symplectic forms for tetrad General Relativity}

Applications of covariant phase space methods to the tetrad action with the bare symplectic potential can be found in \cite{Ashtekar:1999yj,Ashtekar:2008jw,Corichi:2013zza,Corichi:2016zac}. In the rest of this Section, we provide a  comparative analysis of all charges associated with the two choices of symplectic potentials: the bare $\th_{\scr e}$ defined in \Ref{thEC1}, with pre-symplectic 2-form
\be\label{OmEC}
\Om(\d_1,\d_2) = \f12\int_{\Si}\eps_{IJKL} \left(\d_1 \Si^{IJ}\w \d_2 \om^{KL} - \d_2 \Si^{IJ}\w \d_1 \om^{KL} \right).
\ee
And the dressed, metric-equivalent one $\theta^{\prime}_{\scr e} := \theta_{\scr e} + d\alpha\equiv \th_{\scr g}$ defined in \Ref{thg}, with pre-symplectic 2-form
\begin{subequations}
\begin{align} \label{OmEC1}
\Om'(\d_1,\d_2)&= \Om(\d_1,\d_2) + \Om_\a(\d_1,\d_2),\\
 \label{OmAlpha}
\Om_\a(\d_1,\d_2) & = -\f12 \int_{\p\Si} \eps_{IJKL}~ \left[ \d_1\left(\Si^{IJ} e^{\rho K} \right) \delta_2 e^L_\rho  - \d_2\left(\Si^{IJ} e^{\rho K} \right) \delta_1 e^L_\rho\right].
\end{align}
\end{subequations}

To study the charges, we briefly recall that the tetrad Lagrangian \Ref{SEC} has two different gauge symmetries: internal Lorentz transformations
\be\label{Lorentz}
\d_\l e^I :=\l^I{}_J e^J, \qquad \d_\l\om^{IJ} := -d_\om \l^{IJ},
\ee
and diffeomorphisms
\begin{subequations}\label{diffeos}\begin{align}
 \d_\x e^I &:= \pounds_\xi e^I = i_\xi de^I +d(i_\xi e^I)= i_\xi d_\om e^I +d_\om(i_\xi e^I) - (i_\xi \om^I{}_J) e^J,\\
 \d_\xi\om^{IJ}&:=\pounds_\xi \om^{IJ} = i_\xi d\om^{IJ} +d(i_\xi \om^{IJ})=i_\xi F^{IJ} +d_\om(i_\xi \om^{IJ}).
\end{align}\end{subequations}
The Lie derivative appearing here is not gauge-covariant under the internal Lorentz transformations. 
Any linear combination is also a symmetry of the theory, and this fact can be used to define a  covariant Lie derivative
\be\label{defL}
L_\xi := \pounds_\xi + \d_{i_\xi\om},
\ee
whose action on the fields equals \Ref{diffeos} with the last terms removed.

\subsection{Internal Lorentz tranformations}

The tetrad Lagrangian is exactly invariant under internal Lorentz transformations, hence the Noether current coincides with the symplectic potential,
\be\label{jL}
j(\d_\l) = \th_{\scr e}(\d_\l) = -\f12d \left(\eps_{IJKL} \l^{IJ} \Si^{KL}\right),
\ee
and the Noether charge is 
\be\label{QL}
Q_\l :=\int_{\Si}j(\d_\l) =  -\f12\int_{\p\Si}\eps_{IJKL}~ \l^{IJ} \Si^{KL}. 
\ee
For the Hamiltonian charges, a simple calculations gives
\begin{align}\label{HL}
\sd H_\l &\approx -\f12\int_{\Si}\eps_{IJKL} \left(\d \Si^{IJ}\w d_\om \l^{KL} + \d_\l \Si^{IJ}\w \d \om^{KL} \right)\nn\\
& = \d Q_\l - \int_\Si \th_{\scr e}([\d,\d_\l]) = -\f12\int_{\p\Si}\eps_{IJKL} ~\l^{IJ} \d\Si^{KL} .
\end{align}
This expression is integrable for field-independent gauge transformations, $\d \l=0$. In this case the Hamiltonian charges exist, and coincide with the Noether charges \Ref{QL}.
Conversely, the Noether charges are well-defined also for field-dependent gauge transformations, but they are not canonical generators.
We also notice that in the presence of isometries these charges are independent of the integration surface in vacuum, so they satisfy a 3d Gauss law like for YM theory. This follows from the fact that under a Killing diffeomorphism $\xi$, the tetrad undergoes a gauge transformation $\l_\xi^{IJ}=-D^{[I}\xi^{J]}$ (see the end of this Section, Eq.~\eqref{deflxi}), and 
\be\nn
j(\d_{\l_\xi}) = -\f12 \eps_{IJKL} d_\om \l_\xi^{IJ} \w \Si^{KL} = \f12 \eps_{IJKL} i_\xi F^{IJ} \w \Si^{KL} = 
\f13 R^\m{}_\a\xi^\a \eps_{\m\n\r\s}dx^\n\w dx^\r\w dx^\s \approx 0
\ee
in vacuum. These internal Lorentz charges have no counterpart in metric variables.

Let us now compare these results with those obtained using the metric-equivalent, gauge-invariant (pre)-symplectic form \Ref{OmEC1}.
Specializing the contribution of the dressing 2-form to internal gauge transformations, one gets 
\be
j'(\d_\l) = \th'_{\scr e}(\d_\l) \equiv 0,
\ee
with identically vanishing Noether charge. As for the Hamiltonian charge, we have
\be
\Om_\a(\d,\d_\l)=\f12\int_{\p\Si}\eps_{IJKL} ~\l^{IJ} \d\Si^{KL}.
\ee
This cancels exactly the internal Lorentz charges \Ref{HL} produced by the bare potential. Therefore, both Hamiltonian and Noether charges associated to the internal Lorentz gauge and computed with the gauge-invariant pre-symplectic form \Ref{OmEC1} are vanishing,
\begin{align}
H'_\l \equiv 0 \equiv Q'_\l. 
\end{align}

\subsection{Diffeomorphisms}

With the bare symplectic potential, the Noether current for diffeomorphisms reads
\be \label{jdiffEC}
j(\d_\xi) = \th_{\scr e}(\d_\xi)-i_\xi L_{\scr e} \approx \f12d \left(\eps_{IJKL}~ i_\xi \om^{IJ}~ \Si^{KL}\right),
\ee
with Noether charge
\be\label{QdiffEC}
Q_\xi := \int_\Si j(\d_\xi) = \f12\int_{\p\Si}\eps_{IJKL}~ i_\xi \om^{IJ} ~\Si^{KL}.
\ee
For the Hamiltonian charges, one has
\begin{align}\label{HdiffEC}
\sd H_\xi 
&\approx \f12\int_{\p\Si}\eps_{IJKL} \left( i_\xi\om^{IJ}\d\Si^{KL}  - i_\xi \Si^{IJ}\w \d\om^{KL}\right) \nonumber\\
&= \d Q_{\xi} - \int_{\p \Si}\left( i_{\xi}\th_{\scr e}(\d) + \f12 \eps_{IJKL} ~i_{\d \xi}\omega^{IJ} ~\Si^{KL}\right).
\end{align}
As in the metric case, the simplest integrable Hamiltonians are the field-independent diffeomorphisms tangential to $\p\Si$, for which the Hamiltonian charges coincide with the Noether charges. In any case, the Noether and Hamiltonian charges constructed with the bare tetrad potential differ from the metric ones \Ref{Qdiff} and \Ref{IWmetric} recalled in the previous Section. The difference is manifest since \Ref{HdiffEC} is linear in $\xi$, whereas \Ref{IWmetric} is not. Worse, \Ref{QdiffEC} and \Ref{HdiffEC} are not even gauge-invariant, because of the $i_\xi\om$ term. This non-gauge invariance goes of course hand in hand with the presence of non-zero internal Lorentz charges.

To evaluate the dressed symplectic potential, we first compute
\be\label{adxi}
\a(\d_\xi) = -\f12\eps_{IJKL}\Si^{IJ} e^{K\r}\d_\xi e^L_\r = -\f12\eps_{IJKL}\Si^{IJ} (D^K\xi^L+i_\xi\om^{KL}),
\ee
where we see the appearance of
\be\label{ktetrad}
\k_\xi = - \f12 \eps_{IJKL} ~ \Si^{IJ} D^K\xi^L = -\f12 \eps_{\m\n\r\s} \na^\r\xi^\s dx^\m\w dx^\n, 
\ee
the Komar 2-form in tetrad language. Adding the exterior derivative of \Ref{adxi} to \Ref{jdiffEC} we recover the metric result \Ref{jEH} for the Noether charges.
For the Hamiltonian charges, the contribution of the dressing 2-form is
\begin{align}\label{Omadiff}
\Om_\a(\d,\d_\xi) & = -\f12 \eps_{IJKL} \int_{\p\Si} \d(\Si^{IJ} e^{\r K})\pounds_\xi e_\r^L - \pounds_\xi ( \Si^{IJ} e^{\r K}) \d e^L_\r.
\end{align}
There are two useful ways of manipulating this expression. One is
\begin{align}\label{Omadiff1}
\Om_\a(\d,\d_\xi) 
& = -\f12  \eps_{IJKL} \int_{\p\Si}  \d(\Si^{IJ} e^{\r K}\pounds_\xi e_\r^L) - \Si^{IJ} e^{\r K}\pounds_{\d\xi} e_\r^L 
-\int_{\p\Si}\pounds_\xi\a(\d),
\end{align}
leaving the last term implicit. With $\p\Si$ compact, we can replace this term with $i_\xi d\a(\d)$. It then combines with the $i_\xi\th_{\scr e}(\d)$ term in \Ref{HdiffEC} to give the metric symplectic potential. 
For the first term in \Ref{Omadiff1} we use the identity 
\be
e^{\n[I}L_\xi e^{J]}_\n= e^{\n[I}\pounds_\xi e^{J]}_\n - i_\xi\om^{IJ} = D^{[I}\xi^{J]}, 
\ee
which produces the Komar 2-form \Ref{ktetrad}. 
After these manipulations, adding \Ref{Omadiff1} to the bare contribution \Ref{HdiffEC} gives
\begin{align}\label{HdiffEC'}
\sd H'_\xi 
&\approx \int_{\p\Si} \left( \d\k_\xi -i_\xi\th_{\scr g}(\d) - \k_{\d\xi}\right) , 
\end{align}
recovering the metric expression for the charges.

In the second way, we rewrite \Ref{Omadiff} as
\begin{align}
\Om_\a(\d,\d_\xi) 
\label{Omadiff2} 
 & = -\f12 \eps_{IJKL} \int_{\p\Si} \d\Si^{IJ} e^{\r K}\pounds_\xi e_\r^L +\Si^{IJ}  \d e^{\r K}L_\xi e_\r^L -L_\xi ( \Si^{IJ} e^{\r K}) \d e^L_\r.
\end{align}
The first term is the only non-gauge-invariant one, but combines with the $i_\xi\om$ term of \Ref{HdiffEC} to give a manifestly gauge-invariant expression,
\begin{align}\label{Hdiffeo'}
\sd H'_\xi &\approx \f12 \int_{\p\Si} \eps_{IJKL} \Big( L_\xi (\Si^{IJ} e^{\r K}) \d e^L_\r - \d(\Si^{IJ} e^{\r K}) L_\xi e_\r^L  - i_\xi\Si^{IJ}\w\d\om^{KL}\Big).
\end{align}
This equation is the main new result of this Section, and provides an expression in tetrad-connection variables that is fully equivalent to the metric charges \Ref{IWmetric}. Notice also that although we are allowing for field-dependent diffeomorphisms, no explicit variations $\d\xi$ appear in this way of writing the charges.

\subsection{Covariant diffeomorphisms}

By linearity, the expressions for the Noether currents and Hamiltonian charges can be simply added up to deal with the case of arbitrary linear combinations of diffeomorphisms and internal Lorentz transformations. 
For the Noether current, adding up Eqs.~\eqref{jL} and \eqref{jdiffEC} we get
 \be \label{jCov}
j\left(\d_{\left(\l,\xi\right)}\right) = j(\d_\l) + j(\d_\xi) \approx \f12d\left[\eps_{IJKL}\left(i_\xi\om^{IJ}-\l^{IJ}\right)\Si^{KL}\right].
\ee
For Hamiltonian charges, adding up Eqs.~\eqref{HL} and the first line of \Ref{HdiffEC} we get
\be\label{Hgen}
\sd H_{\left(\l,\xi\right)} = \sd H_\l + \sd H_\xi \approx \f12 \int_{\p\Si} \eps_{IJKL} \big[ (i_\xi\om^{IJ}-\l^{IJ})\d\Si^{KL}  - i_\xi \Si^{IJ}\w \d\om^{KL}\big].
\ee

Among the linear combinations, it is useful to look at the covariant Lie derivative \Ref{defL}. The associated quantities can be obtained from the above formulas specializing to $\l=i_\xi\om$. Notice that this is a field-dependent gauge transformation, but our formulas are valid in this case as well. 
We derive, respectively, from Eqs.~\eqref{jCov} and \eqref{Hgen}
\be
j(L_\xi) \approx 0
\ee
and
\be\label{HLxi}
\sd H_{L_\xi} \approx  -\f12\int_{\p\Si}\eps_{IJKL} i_\xi \Si^{IJ}\w \d\om^{KL}.
\ee

To obtain the correspondent expressions computed with the dressed symplectic potential, it suffices to observe that $\sd H_{\l}'=0$, hence
\be
\sd H'_{L_{\xi}} = \sd H'_{\xi}.
\ee
%

\subsection{Isometries and the Kosmann derivative prescription}

Isometries are characterized in metric variables 
by the Killing equations
\be
\pounds_\xi g_{\m\n}=2\na_{(\m}\xi_{\n)}=0, \qquad R_{\s\m\n\r}\xi^\s=\na_\m\na_\n\xi_\r.
\ee
While an isometry leaves the metric invariant, its tetrad can still transform, but  by an internal Lorentz transformation at most. This
means that the covariant Lie derivative associated with a Killing vector is a gauge transformation determined by the vector itself, 
\be\label{eKilling}
L_\xi e^I=\l_\xi{}^I{}_J e^J, \qquad L_\xi\om^{IJ}= -d_\om\l^{IJ}_\xi.
\ee
These conditions are solved by 
\be\label{deflxi}
\l_\xi{}^{IJ} = -e^{\r I}L_\xi e^{J}_\r = -D^{[I}\xi^{J]}.
\ee
Let us now take a linear combination of a diffeomorphism and a fine-tuned gauge transformation
\be
{\cal K}^{(e)}_\xi e^I:=\pounds_\xi e^I + \d_{\bar\l} e^I,
\ee
with field-dependent parameter
\be
\bar\l^{IJ}:=i_\xi\om^{IJ} - \l_\xi^{IJ}.
\ee
This Kosmann derivative \cite{TedMohd,Prabhu:2015vua} can be defined for an arbitrary diffeomorphism (for its extension to tensors and spinors see \cite{Aneesh:2020fcr}), and satisfies by construction ${\cal K}^{(e)}_\xi e^I=0$ for a Killing transformation. Its key property is to reproduce the metric charges from the bare tetrad symplectic potential. This follows from the observation that
\be\label{equivalence}
\th_{\scr e}({\cal K}^{(e)}_\xi) = \th_{\scr e}({\pounds}_\xi) + \th_{\scr e}(\d_{\bar\l} ) \equiv \th_{\scr g}({\pounds}_\xi). 
\ee
From our perspective, this result is easy to understand from the underlying difference between the symplectic potentials. In fact, notice that
\be
\th_{\scr e}(\d_{\bar \l} ) = d\a(\pounds_\xi).
\ee
Therefore, \Ref{equivalence} follows from the more general equivalence \Ref{thda}.

%
%

\section{Time gauge and adapted tetrads} \label{Sec:TG}

Even though the bare tetrad symplectic potential gives rise to non-zero internal Lorentz charges, these charges vanish if one restricts the variations to adapted tetrads only, namely tetrads with one element always aligned with the normal hypersurface. In the case of a space-like hypersurface, we adapt the tetrad taking
\be\label{tg}
e^0 = n, \qquad n^2=-1.
\ee
This partial gauge-fixing  breaks  the boost part of the internal Lorentz transformations, leaving only an SU(2) symmetry acting on the internal indices $i=1,2,3$. It is often referred to as time gauge, and it is typically used in General Relativity with Ashtekar-Barbero variables and in Loop Quantum Gravity \cite{ThiemannBook}.
As a result of \Ref{tg}, the pull-back of the bare symplectic potential simplifies to
\be\label{baretg}
\th_{\scr e}^{\sscr tg}(\d) \stackrel{\Si}{=} \f12\eps_{ijk}~e^i\w e^j\w\d\om^{0k}. 
\ee
It contains only the boost part of the connection, which corresponds to the extrinsic curvature.
It is then easy to see that the internal charges \Ref{HL} are all zero: 
\be\label{split}
\sd H_\l \approx -\f12 \int_{\p\Si}\eps_{IJKL}~ \l^{IJ}\d\Si^{KL} = - \int_{\p\Si}\eps_{ijk} (\l^{0i}\d\Si^{jk} + \l^{jk}\d\Si^{0i} ) = 0.
\ee
The first term vanishes because $\l^{0i}=0$ for the little group SU(2) preserving the gauge-fixed adapted tetrad, and the second term because the pull-backs of 
$n_\m$ and $\d n_\m$ on the hypersurface vanish. Fixing the time gauge has removed all internal charges of the bare potential, there is no more the need to add the dressing 2-form to achieve this.

This however does not mean that the time-gauge bare tetrad symplectic potential fully coincides with the metric one, and in fact, it still doesn't.
One way to see it is to show that the dressing 2-form does not completely vanish with adapted tetrads. Explicitly, one finds
\be \label{alphatg}
\a^{\sscr tg}(\d) = \int_{\p\Si} n^\m \hat r_i \d e_\m^i dS = -\int_{\p\Si} n_\m \hat r_\n \d g^{\m\n} dS.
\ee
We remark that it is now a purely metric expression, in agreement with the fact that in the time gauge we have removed all internal charges.
But \eqref{alphatg} not being zero, we conclude that \Ref{baretg} still differs from the metric symplectic potential, and so the associated phase space. In particular for a diffeomorphism we have 
\be\label{atgd}
\a^{\sscr tg}(\d_\xi) = - \int_{\p\Si} \eps_{ijk}\Si^{jk} (D^{[0}\xi^{i]}+i_\xi \om^{0i}) = 2\int_{\p\Si} n_\m \hat r_\n \na^{(\m}\xi^{\n)} dS.
\ee
Thus, the diffeomorphism charges with the bare, time-gauge symplectic potential differ from the metric expression \Ref{IWmetric} by the term  $\d\a^{\sscr tg}(\d_\xi) -\a^{\sscr tg}([\d,\d_\xi])$,  {giving} 
\be
\sd H_\xi \approx  \int_{\p\Si} \d\k_\xi -i_\xi \th(\d) - \k_{\d\xi} +2 \d (n^\m \hat r^\n \na_{(\m}\xi_{\n)} dS) -2 n^\m \hat r^\n \na_{(\m}\d\xi_{\n)} dS.
\ee
Inspection of \Ref{atgd} shows that this term vanishes for a Killing vector.
Therefore, the bare time-gauge tetrad symplectic potential gives the same Killing charges as the metric theory.

The same considerations apply to any hypersurface, not just space-like ones. 
For a time-like hypersurface with $n$ space-like, one can adapt say $e^3=n$, and the little group is SU(1,1). For a null hypersurface with $n$ null, one can adapt say $(e^0+e^3)/\sqrt{2}=n$, and the little group is ISO(2).
In both cases, all SU(1,1) and ISO(2) charges are zero for the same argument \Ref{split}. 

One final comment to connect with some literature \cite{Freidel:2015gpa,Freidel:2016bxd,Geiller:2017whh,Freidel:2019ees}. Let us look back at \Ref{baretg}, and add and subtract the quantity
\be
\b\left( \f12\eps_{ijk}e^i\w e^j\w\d\om^{mn} \eps^k{}_{mn} \right)\equiv \b d(e_i\w\d e^i).
\ee
This gives
\be
\th_{\scr e}^{\rm tg}(\d) = \f12\eps_{ijk}e^i\w e^j\w\d A^k -\b  d(e_i\w\d e^i),
\ee
where  $A^k:=\om^{0k}+\b\eps^k{}_{mn}\om^{mn}$ is the Ashtekar-Barbero connection. Written in this way, the gauge-fixed bare symplectic potential has a bulk and a boundary contributions, respectively, in $\d A$ and in $\d e$. It can be easily shown that each term individually has non-vanishing SU(2) charges
\be
\sd H^{\rm tg}_{\l} = \int_{\p\Si} \l_{ij} \d\Si^{ij},
\ee
equal and opposite in sign,  in agreement with the total charge being zero.
This manipulation, which is quite natural from the LQG viewpoint, was pointed out in \cite{Engle:2010kt,Freidel:2015gpa}, where an interpretational splitting between the first and second terms as bulk and boundary degrees of freedom was proposed, and this has given rise to subsequent work on edge modes \cite{Freidel:2016bxd,Cattaneo:2016zsq,Geiller:2017whh,Freidel:2019ees}.

\section{First-order Lagrangians and the Barbero-Immirzi parameter} \label{Sec:Immirzi}

The tetrad Lagrangian is often written in the first-order formalism, with an independent spin connection and only first derivatives appearing:
\begin{align}\label{SEC1}
& L_{\scr (e,\om)} = 
\f12 \eps_{IJKL}~e^I\w e^J\w F^{KL}(\om) - 2\L\eps.
\end{align}
In the absence of matter couplings sourcing torsion, this Lagrangian is equivalent to \Ref{SEC}. 
In the paper so far we have considered second-order Lagrangians, and we stress that the differences in bare symplectic potentials and their consequences follow from the use of the tetrad instead of the metric, and not from a switch from second to first order which often accompanies the use of tetrads.
We report in this Section the difference between tetrad and metric symplectic potentials when using a first-order formalism.
The results are very similar. One aspect worth mentioning  is that the potentials now differ off-shell by a bulk term, and not just by an exact 3-form. This off-shell difference is a simple consequence of the fact that although the connection field equations are equivalent in the two choices of variables, they are off-shell different.

We consider a first-order formalism with connections that can a priori carry torsion, but are still metric or tetrad compatible, namely $\na_\m g_{\n\r}=0=\om_\m^{(IJ)}$. 
This generalization affects our previous manipulations in two aspects. First, the curvature of $\G$ has 36 independent components, not just 20 (see, \emph{e.g.}, \cite{Speziale:2018cvy} for a decomposition into its 6 irreps) and it is only one part of the commutator: 
\begin{subequations}
\begin{align}
& [\na_\r,\na_\s]f^\m = R^\m{}_{\n\r\s} f^\n - T^\n{}_{\r\s}\na_\n f^\m,
\\ & R^\m{}_{\n\r\s}(\G) = 2(\p_{[\r}\G^\m_{\s]\n}+\G^\m_{[\r|\l}\G^\l_{\s]\n}), 
\end{align}
\end{subequations}
where $T^\m{}_{\n\r} = 2\G^\m_{[\n\r]}$ is the torsion.
Second, the familiar rule to pass from a covariant divergence to a boundary term through Stoke's theorem leaves a bulk term behind:
\be\label{divtor}
\sqrt{-g}\na_\m v^\m = \p_\m(\sqrt{-g}v^\m) + \sqrt{-g} \, T^\m{}_{\m\n}v^\n.
\ee

Let us now look at the Lagragian \Ref{SEC1}. 
In the first-order formalism with independent connection variables, there is a second dimension-2 term in the Lagrangian, whose coupling constant is (inversely) proportional to the Barbero-Immirzi parameter $\g$. We include this term in our analysis for completeness, but the reader interested only in the basic Lagrangian can easily deduce the relevant formulas setting $1/\g=0$.
The more general tetrad Lagrangian reads
\begin{subequations}
\begin{align}\label{SEC2}
& L_{\scr (e,\om,\g)} = 
P_{IJKL}~e^I\w e^J\w F^{KL}(\om) - 2\L\eps, \\ &P_{IJKL} = \f12\eps_{IJKL} + \f1\g\eta_{I[K}\eta_{L]J}.
\end{align}
\end{subequations}
Its variation gives
\begin{align}
\d L_{\scr (e,\om,\g)} &= 
\d e^I\w \FE{e}_I +\d \om^{IJ}\w  \FE{\om}_{IJ}   +d\th_{\scr (e,\om,\g)},
\end{align}
with field equations
\begin{subequations} \label{FE1st}\begin{align}
& \FE{e}_I =  2P_{IJKL} ~e^J\w \left(F^{KL}-\f23\L ~e^K\w e^L \right), \\ 
& \FE{\om}_{IJ}= -2 P_{IJKL} ~e^K\w d_\om e^L,\label{FE2}
\end{align}
\end{subequations}
and bare symplectic potential 
\be\label{thECH}
\th_{\scr (e,\om,\g)}(\d):= P_{IJKL}~e^I\w e^J\w \d\om^{KL}. 
\ee

In metric and affine connection the equivalent Lagrangian reads 
\be\label{EHg}
L_{(g,\G,\g)} = \left(g^{\m\n}R_{\m\n}(\G)-2\L \right)\eps - \f1{2\g} \tl\eps^{\m\n\r\s}R_{\m\n\r\s}(\G)d^4x.
\ee
 To take the variation, we use the identity
 \be
 \d R^\m{}_{\n\r\s}(\G) = 2 \left(\na_{[\r}\d\G^\m_{\s]\n}+\G^\l_{[\r\s]}\d\G^\m_{\l\n}\right),
 \ee
 from which it follows that
\be
\d L_{(g,\G,\g)} = \Big[\FE{g}{}_{\m\n} \d g^{\m\n}  +\FE{\G}{}_{\m}^{\n\r} \d \G^\m_{\n\r} \Big]\eps + d\th_{(g,\G,\g)},
\ee
with field equations
\begin{subequations}
 \begin{align}
& \FE{g}{}_{\m\n} = G_{\m\n}+\L g_{\m\n} +\f1\g\eps_{(\m}{}^{\l\r\s} R_{\n)\l\r\s} \\
& \FE{\G}{}_{\m}^{\n\r} =g_{\m\l} \left(T^{\n,\l\r}+2T^\a{}_{\a}{}^{[\l} g^{\r]\n}\right) -\f1{2\g}\left(T^\n{}_{\a\b} \eps_\m{}^{\r\a\b} + 2T^\a{}_{\a\b}\eps_\m{}^{\n\r\b}\right)\label{GFE1}
\end{align}
\end{subequations}
and bare symplectic potential
 \begin{align}
& \th_{(g,\G,\g)\n\r\s}= \th_{(g,\G,\g)}^\m \eps_{\m\n\r\s}, \qquad \th_{(g,\G,\g)}^\m = \left(2g^{\r[\m}g^{\n]\s} -\f1\g \eps^{\m\n\r\s}\right) g_{\r\l} \d \G^\l_{\n\s}.\label{thetagG}
\end{align}
For further details on the first order formalism see for instance \cite{Hehl:1976kj,DeLorenzo:2018odq,Chakraborty:2018qew}.

To compare the two symplectic potentials we proceed as in the second-order theory, since the relation  \Ref{defom} holds also in the presence of torsion.
This time, we find
\begin{align}\nn
\th^\m_{(e,\om,\g)} 
&= 2 g^{\r[\n} \d \G^{\m]}_{\n\r} + \na_{\n}\left(2e^{[\n}_I  \d e^{\m]I}\right) +\f1\g\eps^{\m\n\r\s}\left(g_{\n\l}\d\G^\l_{\r\s} +\na_\s(e_{I\n}\d e^I_\r) \right)\\\nn
& = \th^\m_{(g,\G,\g)} + \f1 e\p_\s\left(2e e^{[\s}_I  \d e^{\m]I} +\f1\g\tl\eps^{\m\n\r\s}e_{I\n}\d e^I_\r\right) \\
& \quad +T^\m{}_{\n\r}e^\n_I\d e^{I\r} + \f1{2\g}\eps^{\m\n\r\s}T^\l{}_{\n\r} \left(\d e_{I\l}e^I_\s - e_{I\l}\d e_\s^I\right),
\label{thetas}
\end{align}
with the additional torsion-dependent bulk piece due to \Ref{divtor}.
The relation \Ref{thetas} can be rewritten in terms of forms as
\be\label{thdaT}
\th_{(g,\G,\g)}  = \th_{(e,\om,\g)}+ d\a + {\cal  T},
\ee
where
\be
\a(\d) := \star (e_{I}\w \d e^{I}) +\f1\g e_{I}\w \d e^{I} = -P_{IJKL}~e^I\w e^J \left(e^{\r K}\d e^L_\r\right),
\ee
and the torsion bulk piece reads as
\be
{\cal T}_{\a\b\g} = \f1{3!}\eps_{\a\b\g\m} T^\m{}_{\n\r}e^\n_I\d e^{I\r} dx^\a \w dx^\b\w dx^\g + 
\f1{\g} (T_I\w \d e^I - \d T_I\w e^I - e_{I\l}\d T^\l\w e^I). 
\ee
See  Appendix~\ref{AppT1} for more details.
We see that in the presence of torsion, the bare symplectic potentials differ not just by an exact form, but also by a bulk term. The reason for this is that 
\Ref{FE2} is not equal to \Ref{GFE1}, but differs by a boundary term. This difference vanishes on-shell, and corresponds to the fact that the affine and Lorentz connections have different ways to encode torsion.

Having kept  the short-hand notation $\d\om$ in most of our previous formulas, we can adapt most of them easily to the case when $\om$ is arbitrary. 
For the geometric decomposition, the curvature part contains a torsion piece; see Eq.~\Ref{cusa2}. We have
\begin{align} 
\Th_{\scr (e,\om)}(\d) =& s \int_\Sigma \eps_{IJKL}\big[-\d \big(\Si^{IJ} \w n^{K} d_\om n^L \big) + \d \Si^{IJ} \w n^{K} d_\om n^L - 2 T^I\w e^J \, n^K  \d n^L  \big] \nn\\ 
& +s\int_{\p\Si}\eps_{IJKL} ~\Si^{IJ} n^K \d n^L  
\end{align}
and
\begin{align}
\Th_{\scr (e,\om)}(\d) =&\int_{\cal N}\eps_{IJKL} \big[ -\d\big(\Si^{IJ} \w l^{K}d_{\omega}n^L \big)+ \d \Sigma^{IJ} \w l^K d_\om n^L + \Sigma^{IJ}\w \left(\d l^{K}d_\om n^L + \d n^{K} d_{\om}l^L \right)\big] \nn\\
&- \int_{\cal N} 2\eps_{IJKL} T^{I}\w e^{J} l^{K}\d n^{L} + \int_{\p \cal N}\eps_{IJKL}\Sigma^{IJ} l^{K} \d n^L.
\end{align}
The part in $\g$ gives the contribution
\be
e_I\w e_J \w\d\om^{IJ} = T_I\w \d e^I - \d T_I\w e^I - d(e_I\w \d e^I).
\ee
Adding \Ref{thdaT} to the two above symplectic potentials, we obtain the corresponding formulas for the geometric decomposition in metric-connection variables.

\subsection{Surface charges}
All charges, Noether and Hamiltonian, with this bare symplectic potential, can be deduced from the ones computed above with the trivial replacement 
\begin{equation}\label{BIsub}
\f12\eps_{IJKL} \mapsto P_{IJKL},
\end{equation}
namely
\begin{subequations}
\begin{align} 
\label{HgenP}  
\sd H_{(\l,\xi)} &= \int_{\p\Si} P_{IJKL} \left[ \left(i_\xi\om^{IJ}-\l^{IJ}\right)\d\Si^{KL}  - i_\xi \Si^{IJ}\w \d\om^{KL}\right], \\  
 \sd H_{L_\xi} &=  - \int_{\p\Si} P_{IJKL}~ i_\xi \Si^{IJ}\w \d\om^{KL}.
\end{align}
\end{subequations}
The corresponding formulas for the Noether charges can be obtained with the same replacement.
The explicit dependence of the charges on $\g$, even in the absence of torsion, is one more peculiarity of using the bare tetrad symplectic potential. There is in fact no such contribution when using metric variables, even in the first-order theory with the Lagrangian \Ref{EHg}: the contribution to the diffeomorphism charges proportional to $1/\g$ is an exact 2-form, and vanishes in the customary case of compact (but not necessarily connected) surfaces. The independence from $\g$ is a natural feature, since the physical solutions don't depend on $\g$ in the absence of torsion.

For the dependence of the charges on higher-order topological terms see \cite{Corichi:2016zac,Frodden:2017qwh,Godazgar:2020gqd,Godazgar:2020kqd}. 
 {For further discussions on the role of torsion in computing charges and the first law, see e.g.} \cite{DeLorenzo:2018odq,Chakraborty:2018qew,Aneesh:2020fcr,Gallegos:2020otk}.

\section{Cohomological methods and Barnich-Brandt charges} \label{Cohomological method}

There is an alternative definition of surface charges that avoids both ambiguities $I$ and $II$. 
It is  based on ideas of Anderson and Torre \cite{Anderson:1996sc}, and developed by Barnich, Brandt and Henneaux \cite{Barnich:2000zw,Barnich:2001jy}; see  \cite{Compere:2018aar} for a recent review. 

The idea is to work directly with the field equations, rather than with the Lagrangian.
Let us look back at Eq.~\Ref{dL}, and focus on the term containing the field equations. 
We can split this term into a piece linear in the gauge parameters $\eps$, and a piece containing their derivatives,
\be\label{Edphi}
\FE{\phi} \d_\eps\phi = N(\eps) + dS(\eps).
\ee
On general grounds, the two terms on the right-hand side are, respectively, the Noether identities -- which vanish exactly -- and (once pulled-back on a hypersurface) the constraints generating the symmetry -- which vanish on-shell. The idea is to use the term with the constraints to define a \emph{weakly-vanishing} Noether current $S(\eps)$.
Comparison with \Ref{jW} shows that it differs from Wald's definition of the Noether current by at most an exact form and a total variation. 
The weakly-vanishing Noether current is free from the ambiguity $I$ because it is constructed from the field equations and not from the Lagrangian. As for the ambiguity $II$, there is a priori still a cohomology ambiguity when extracting $S(\eps)$ from \Ref{Edphi}. However, this ambiguity is eliminated with the prescription of taking the unique weakly-vanishing 3-form.

To fix ideas with an example, for General Relativity in metric variables, one has 
\begin{align}
 \FE{g}  \d_\xi g &= (G_{\m\n}+\L g_{\m\n})\pounds_\xi g^{\m\n} \sqrt{-g}d^4x \nn\\&=
2 \xi_\n \na_\m G^{\m\n}  \sqrt{-g} d^4x -\p_\m(2\sqrt{-g} (G^{\m\n} +\L g^{\m\n})\xi_\n)d^4x.
\end{align}
This identifies $N_{\scr g}(\xi):= 2\xi_\n \na_\m G^{\m\n} \eps $ are 
the Bianchi identities, \emph{i.e.}, the Noether identities for diffeomorphisms. The second term is the weakly-vanishing Noether current, which can be compactly written as the 3-form
\be\label{Sg}
S_{\scr g}(\xi)=  \FE{e}_I\xi^I = \f13  \eps_{\a\b\g\m} (G^{\m\n}+\L g^{\m\n}) \xi_\n \, dx^\a\w dx^\b\w dx^\g \approx 0,
\ee
where we used \Ref{EEe}. 
Its pull-back on a hypersurface gives the Hamiltonian constraints contracted with $\xi^\m$.
 Comparing with \Ref{jEH} defined earlier, we see that we have picked the representative in the cohomology class of Noether currents that vanishes on shell.

To obtain the Hamiltonian generators from the weakly-vanishing Noether current, 
one can use a method based on the homotopy operator of Anderson. This is a  map from spacetime $p$-form to $(p-1)$-forms, and we refer the reader to \cite{Anderson:1996sc,Barnich:2001jy} for its formal definition and properties. We only need here its action on a top 4-form and on a 3-form, which are given by
\begin{subequations}\label{homotopy operator}
\begin{align}
\mathcal{I}^{(4)}_\d&=  \left[\d\phi\f{\d}{\d\p_\m\phi} - \d\phi \p_\n\f{\d}{\d\p_\m\p_\n\phi} + \p_\n\d\phi \f{\d}{\d\p_\m\p_\n\phi} +\ldots  \right] i_{\p_\m}, \\
\mathcal{I}^{(3)}_\d&= \left[\f12\d\phi\f{\d}{\d\p_\m\phi} - \f13\d\phi \p_\n\f{\d}{\d\p_\m\p_\n\phi} + \f23\p_\n\d\phi \f{\d}{\d\p_\m\p_\n\phi} +\ldots  \right] i_{\p_\m}.
\end{align}
\end{subequations}
It can then be explicitly checked, \emph{e.g.} \cite{Compere:2009dp}, that $\th(\d)=\mathcal{I}^{(4)}_\d L$, and that 
\be\label{omWE}
\om(\d_1,\d_2) = W(\d_1,\d_2)+d{\cal E}(\d_1,\d_2),
\ee
where 
\begin{align} \label{defWE}
& W(\d_1,\d_2)=  \mathcal{I}^{(4)}_{\d_1}(\FE{\phi}\w\d_2\phi),
\qquad {\cal E}(\d_1,\d_2)= 
\f12 \mathcal{I}^{(3)}_{\d_1}\th(\d_2).
\end{align}
If we specialize the second variation to a gauge transformation, with the first arbitrary, we find 
\be
W(\d,\d_\eps)=\mathcal{I}^{(4)}_{\d}(\FE{\phi}\w\d_\eps\phi) =  \mathcal{I}^{(4)}_{\d} dS(\eps) =  d\mathcal{I}^{(3)}_{\d}S(\eps),
\ee
where in the last step we used a cohomological property of the homotopy operator.
From this and \Ref{omWE} it follows that 
\be\label{Hcoho}
\sd H_\eps = \int_{\p\Si} \mathcal{I}^{(3)}_{\d}S(\eps) + \int_{\p \Si} \f12 \mathcal{I}^{(3)}_{\d}\th(\d_\eps).
\ee

This expression provides an alternative derivation of the Hamiltonian charges, and shows the ambiguities of covariant phase space methods in a different light. 
First, even though \Ref{Hcoho} depends on $\th$, the action of the homotopy operator in $\cal E$ has a kernel for total variations,  thus explaining the $I$-invariance of the Hamiltonian charges. As for the cohomology ambiguity of type $II$, this has been fixed in $S$ as explained earlier, and therefore it comes entirely from $\th$ in the second term.
This prompts the alternative definition of surface charges where the second term is dropped,
\be\label{BBcharge}
\sd Q^{\scr BB}_\eps:=  \int_{\p\Si} \mathcal{I}^{(3)}_{\d}S(\eps) \equiv \sd H_\eps -  \int_{\p \Si}\f12 \mathcal{I}^{(3)}_{\d}\th(\d_\eps).
\ee
These Barnich-Brandt (BB) surface charges are completely unambiguous. As a price to pay, they differ in general from the Hamiltonian generators.
However,  the difference vanishes in the case of isometries, namely for diffeomorphisms that are Killing and for parallel gauge transformations (namely covariantly constant). In fact, an explicit calculation in General Relativity shows that   
\be\label{Ediffeo}
\mathcal{I}^{(3)}_{\d}\th(\d_\xi) = \eps_{\m\n\r\s} g^{\r\a}\d g_{\a\b} \na^{(\s}\xi^{\b)}  dx^\m\w dx^\n.
\ee
The BB charge for diffeos is thus given by \Ref{IWmetric} plus \Ref{Ediffeo} and the two coincide in the case of a Killing isometry.
Similarly for Yang-Mills theory, one has
\be\label{Egauge}
\mathcal{I}^{(3)}_{\d}\th_{\scr A}(\d_\l)  = \star ( d_A\l \w \d A),
\ee
so the BB charge is \Ref{Hym} plus \Ref{Egauge}, and the two coincide for parallel transported gauge parameters, $d_A\l=0$.

This coincidence in the presence of isometries is very important for the validity of this prescription, since it implies that the BB charges reproduce the usual first law of black hole mechanics as well as the charges associated with asymptotic symmetries. 
On the other hand, they will differ in the study of  edge modes and may have different integrability properties. 

We close this brief review with two remarks that are useful for the applications to tetrad General Relativity below.
First, as we have briefly mentioned, there is a priori a cohomology ambiguity in extracting the current $S(\eps)$ from the field equations \Ref{Edphi}. It is the prescription to pick the weakly-vanishing representative in the equivalence class that eliminates it. One could have taken the same prescription also in the definition of Noether current from the symplectic potential \Ref{jW}, and the two definitions would then match.
Conversely, if one removes this prescription, the cohomology ambiguity in $S(\eps)$ can be used to match the BB charges to the Hamiltonian charges. Indeed, redefining 
\be
S'(\xi):= S(\xi) +\f12 \mathcal{I}^{(4)}_{\d_\xi}L.
\ee
the cohomology methods reproduce the Hamiltonian charges exactly.

The second remark is that the term $\cal E$ defined in \Ref{defWE} vanishes for first-order theories: for these, the symplectic potential does not contain derivatives of the fields, hence it lies in the kernel of  $I^{(3)}_\d$. This means that the BB charges \emph{are different for the same theory whether in the first or second order formalism}. They coincide only in the case of isometries. This point is certainly appreciated in the cohomological literature, where the equivalence is always stated for isometries; see, \emph{e.g.}, \cite{Barnich:2016rwk}. It is however a significative difference from Hamiltonian charges constructed from the (pre)-symplectic 2-form, which always coincide between second order and first order theories.

We now discuss the applications of this method to tetrad GR.

\subsection{Second-order tetrad gravity}

For the tetrad Lagrangian \Ref{SEC} in second-order formalism, with $\om^{IJ}\equiv\om^{IJ}(e)$ the Levi-Civita spin connection, \Ref{Edphi} gives
\begin{align}
\d_{(\l,\xi)} e^I  \w \FE{e}_I&=  N(\lambda, \xi) + dS(\lambda, \xi), 
\end{align}
with Noether identities and weakly-vanishing current are, respectively,
\begin{subequations}
\begin{align}
N(\l,\xi)&= (i_\xi\om^{IJ}-\l^{IJ}) \FE{e}_I\w e_J - i_\xi e^I d_\om \FE{e}_I, \\ 
S(\l,\xi) &:= i_\xi e^I\FE{e}_I. \label{Se} 
\end{align}
\end{subequations}

Notice that the weakly-vanishing Noether current \emph{does not see the internal gauge transformations}. We believe that the reason for this is that the constraint associated to it, part of the torsionless condition, is an identity in the second-order formalism. 
Furthermore, \Ref{Se}  is identical to the metric current given by \Ref{Sg}.
Since the homotopy operator is invariant under field redefinitions, 
we conclude that the BB charges for tetrad gravity in the second-order formalism are the same as the metric ones:
\be
\sd Q_\xi^{\scr BBe} \equiv \sd Q_\xi^{\scr BBg}, \qquad \sd Q_\l^{\scr BBe} \equiv 0.
\ee
This result can be confirmed by a rather lengthy calculation.

These results show that the cohomological prescription gives the same charges in both tetrad and metric variables, unlike the Hamiltonian prescription, \emph{provided one uses a second-order formulation}. The situation changes with a first-order formulation, as we discuss next.

\subsection{First-order tetrad gravity}

Starting from the field equations \Ref{FE1st} and considering a general gauge  variation (internal plus diffeomorphism), we have
\begin{align}\label{NdS}
\d_{(\lambda, \xi)} e^I\w \FE{e}_I + \d_{(\lambda, \xi)}\om^{IJ}\w \FE{\om}_{IJ}   = N(\lambda, \xi) + dS(\lambda, \xi),
\end{align}
with
\begin{align}
N(\l,\xi)&= (i_\xi\om^{IJ} - \l^{IJ}) (\FE{e}_I\w e_J - d_\om \FE{\om}_{IJ}) - i_\xi e^I d_\om \FE{e}_I +i_\xi T^I\w\FE{e}_I +i_\xi F^{IJ}\w\FE{\om}_{IJ}\nn \\
&= \eps_{IJKL} [(\l^{IJ}-i_\xi\om^{IJ}) e^K\w (d_\om T^L - F^{LM} \w e_{M}) + i_\xi e^I e^J\w d_{\om} F^{KL}],
\end{align}
and
\begin{align} \label{weakNCApp}
S(\l,\xi) &:= i_\xi e^I\FE{e}_I + (i_\xi\om^{IJ} - \l^{IJ})  \FE{\om}_{IJ} \nn\\
& = \eps_{IJKL}\big[ i_\xi e^I e^J\w F^{KL} + (i_\xi\om^{IJ} - \l^{IJ}) e^K\w T^L)\big].
\end{align}
See, \emph{e.g}, \cite{Hehl:1985vi,Hehl:1994ue,Barnich:2016rwk,Bonder:2018mfz} for more details.
The weakly-vanishing Noether current now sees both gauge transformations.
To compute the associated charges, we apply \Ref{BBcharge}, finding 
\be \label{QBB1st}
\sd Q^{BB}_{(\l,\xi)} = \int_\Si I^{(3)}_\d S(\l,\xi)
= \f12\int_{\p\Si}\eps_{IJKL} [(i_\xi\om^{IJ}-\l^{IJ})\d \Si^{KL} - i_\xi\Si^{IJ}\w\d\om^{KL}].
\ee
It should be stressed that the cohomological method is significantly simpler to apply in the case of first order theories, as a glance at \Ref{homotopy operator} immediately shows.
The result coincides with the Hamiltonian charges \Ref{Hgen} computed with the bare tetrad symplectic potential, but not with the metric charges.

This fact has two implications. First, the BB prescription gives different charges for the same theory whether in first-order or second-order formulation. Only in the case of isometries, one can recover the same charges using the fine-tuning of \cite{TedMohd,Prabhu:2015vua}, as shown in \cite{Barnich:2016rwk}. This is different from Hamiltonian charges that always give the same answer for any gauge transformation. 
Second, it means that if one works with the first-order formalism, also with the BB charges one runs into the same problem of assigning non-vanishing Lorentz charges to solutions which are in one-to-one correspondence with torsion-less metric General Relativity. 
However, there is now no way out, as long as one sticks with the unique prescription of the weakly-vanishing Noether current. 
An alternative possibility is to give up the uniqueness of the weakly-vanishing current, exploiting the cohomology ambiguity in the definition of $S$, and dress the current with an exact 3-form, 
\be
S'(\l,\xi):=S(\l,\xi)+d\a(\l,\xi),
\ee
constructed so that 
BB charges associated to internal Lorentz transformations vanish, and the diffeomorphism ones coincide with the metric theory.
It is easy to check that this can indeed be achieved, and the exact 3-form to be added is the same DPS  $\alpha(\d)$ in \Ref{DPS}.
Indeed, 
\be
\slashed{\delta}Q^{BB}_{\alpha} = -\mathcal{I}^{(3)} d\alpha(\d_{\lambda}) = \f12 \eps_{IJKL}~\d\left(e^I \w e^J \right)\lambda^{KL} = -\slashed{\delta}Q^{BB}_{\l}.
\ee
And similarly for diffeomorphisms.

\subsection{First-order tetrad gravity with Barbero-Immirzi parameter}

This is a trivial extension that can again be obtained through the substitution \eqref{BIsub}, giving
\be
\sd Q^{BB}_{(\l,\xi)} = \int_{\p\Si}P_{IJKL} [(i_\xi\om^{IJ}-\l^{IJ})~ \d \Si^{KL} - i_\xi\Si^{IJ}\w\d\om^{KL}].
\ee
The result coincides with the Hamiltonian charges \Ref{HgenP} computed with the bare symplectic potential.
We remark the non-trivial dependence of the charges on the Barbero-Immirzi parameter, which has no classical meaning in the absence of torsion. This is, as before, one more reason to doubt the physical meaning of these charges, at least at the classical level.

\subsection{Yang-Mills BB charges in second-order and first-order formalisms}

In concluding this Section we provide a second example showing how the BB charges differ whether one considers a first- or second-order Lagrangian beyond the case of isometry. Applying the cohomological prescription to the second-order YM Lagrangian one gets
\be
\sd Q^{BB, 2\text{nd}}_{\l} = \int_{\p\Si} \tr\left[\l\star\d F -\f12 \star(d_A\l\w\d A)\right],
\ee
where we recognize the general structure \Ref{BBcharge} with the Hamiltonian charge \Ref{Hym} and the additional term \Ref{Egauge}.
YM theory can also be formulated in the first-order formalism using the Lagrangian
\be
L_{\scr A,B} = \tr \left[B \w F - \f12 B\w \star B\right].
\ee
Applying the cohomological prescription in this case gives
\be
\sd Q^{BB, 1\text{st}}_{\l} = \int_{\p\Si} \tr \left[\l \d B\right] \approx \int_{\p\Si} \tr \left[\l \star \d F \right].
\ee
As anticipated, the charges are equal only for isometries.

\section{Conclusions} 

In this paper we investigated the boundary variation of the gravitational Lagrangian in tetrad variables, using either a second-order or first-order formalism, and compared it with the variation in metric variables. Our analysis contains two rather independent parts. In the first part, we studied the geometric decomposition of the boundary variation, a calculation that has applications to the identification of canonical pairs and to the variational problem. We showed that using tetrads one can reproduce the results already known in the literature in an elegant and shorter way, and gain better control in the trickier case of null hypersurfaces. Our results, in particular for null hypersurfaces, allow to bridge among various analysis present in the literature, explaining the relation between different parametrizations and hypothesis used. We highlighted the role of the Bondi gauge, and provided formulas for arbitrary variations that do not require the Bondi gauge. These expose a spin-1 pair whose momentum is the rotational 1-form of isolated horizons.
The main new result in the first part is the derivation of the formulas \Ref{thECnull1} and \Ref{thEHn}, and their relation to various cases studied in the literature. 
The material presented shows, in our opinion, that tetrad variables are a more convenient tool than the metric to address the variational problem and evaluate boundary and corner terms, as already advocated in \cite{Jubb:2016qzt}.

In the second part, we studied the construction of covariant surface charges, first as Noether charges \`a la Wald and as Hamiltonian generators, then as BB charges. Here the use of tetrad variables is not straightforward and requires some care, because a blind application of covariant phase space methods to the tetrad Lagrangian produces results which are generally different from those obtained in metric variables. The difference is traced back to the mismatch of the bare symplectic potentials by an exact 3-form. The difference vanishes for Killing diffeomorphisms, provided one restricts attention to adapted tetrads, and this is the reason why it was not observed in \cite{Ashtekar:2008jw}. Alternatively, the metric Killing charges can also be reproduced without adapting the tetrad, but adding a fine-tuned internal Lorentz transformation,  as done in \cite{TedMohd,Prabhu:2015vua,Barnich:2016rwk}. 
For diffeomorphisms that are not Killing however, the mismatch cannot be avoided adapting the tetrad, 
and this can be physically relevant for balance laws and in the context of charges associated with finite boundaries or subleading corrections to asymptotic symmetries. Mathematically, it can also affect the question of integrability.
If one wants to match all diffeomorphism charges, one 
possibility is to dress the symplectic potential in a suitable way, a procedure which is equivalent to -- but more general than -- the fine tuning of \cite{TedMohd,Prabhu:2015vua,Barnich:2016rwk}, and allows one to work always with ordinary Lie derivative without the need of the Kosmann derivative. This can be done adding the exact 3-form identified in \cite{DePaoli:2018erh}, as explored in this paper, or adding a spinorial boundary action as in \cite{Wieland:2017zkf,Wieland:2019hkz}.
A first consequence of doing so is that one restores the first law of black hole mechanics as a manifestation of the invariance of the Lagrangian under standard diffeomorphisms, as in the metric theory, and further proves its invariance under cohomological ambiguities \cite{DePaoli:2018erh}.
The fact that the spinorial boundary action of \cite{Wieland:2017zkf,Wieland:2019hkz} reproduces the same results of the dressing form calls out for a closer comparison of the two approaches, which we leave for future work. 
Among the results presented in the second part we highlight the new covariant expression for the diffeomorphism charges in tetrad variables given by \Ref{Hdiffeo'}.

The situation is somehow orthogonal if one uses the BB prescription for the charges: these are the same in tetrad or metric variables, however differ if one uses a first-order or second-order Lagrangian, unless they refer to Killing diffeomorphisms, or covariantly constant gauge parameters, in which case they coincide. This situation is not particular to General Relativity, but it is an inherent structure of the cohomological methods used.

We hope that the results and comparisons presented will help set the use of surface charges with tetrad variables on firmer grounds, and be taken as a starting point to address in these variables open questions currently being explored in both metric or tetrad variables, like renormalization of the symplectic structure and charges, multipole expansions, edge modes and entanglement entropy.

In the course of the paper we have pointed out a few reasons to prefer a prescription for charges in tetrad variables that match those of metric variables, and used the dressing 2-form to achieve it. 
The same exact 3-form arises naturally in the geometric approach of \cite{Gomes:2018shn}. It would be interesting to see whether it will appear also in the Batalin-Vilkovisky approach currently being developed for the Einstein-Cartan action \cite{Canepa:2020ujx}.
The question of what is the right prescription will likely require additional considerations. Our reasons were purely classical, like the fact that the covariant phase space is a structure associated to the space of solutions, and these are in one-to-one correspondence between the metric and tetrad formulation of theory, or their contrasting dependence on $\g$. However, tetrad variables are argued to provide a preferable path towards quantum gravity, and it may be that the additional internal Lorentz charges are given physical weight in a quantum context, as argued for instance in \cite{Freidel:2015gpa}. On the other hand, it has been shown that a certain quantization of the area \`a la LQG arises also in a context with vanishing internal Lorentz charges \cite{Wieland:2017cmf}. We leave further investigations of these issues to future work.

\subsection*{Acknowledgments}
Si.S. is grateful to Abhay Ashtekar, Wolfgang Wieland and Marc Geiller for many precious discussions on surface charges. R.O. thanks Geoffrey Comp\`ere and Ali Seraj for interesting discussions on covariant phase space methods, and Ernesto Frodden and Diego Hidalgo for discussions on the first-order gravity.
R.O. also thanks the CPT for funding his visit in November 2018 and for the hospitality during his stay in March 2019, supported by the COST Action GWverse CA16104 under the Short Term Scientific Missions programme.
R.O. is funded by the European Structural and Investment Funds (ESIF) and the Czech Ministry of Education, Youth and Sports (MSMT) (Project CoGraDS - CZ.02.1.01/0.0/0.0/15003/0000437).


\appendix
\section{Conventions}\label{AppN}

We denote by $\ut{\eps}_{\m\n\r\s}$ the completely antisymmetric spacetime density with $\ut{\eps}_{0123}=1$, and 
$\tl\eps^{\m\n\r\s}\ut{\eps}_{\m\n\r\s}=-4!$. It is related to the volume 4-form by
\be
\eps:=\f1{4!}\eps_{\m\n\r\s}~dx^\m\w dx^\n\w dx^\r\w dx^\s, \qquad \eps_{\m\n\r\s}:=\sqrt{-g} \, \ut{\eps}_{\m\n\r\s}.
\ee
For the internal Levi-Civita $\eps_{IJKL}$ the density notation is unnecessary, and we use the same convention, ${\eps}_{0123}=1$. Hence the tetrad determinant is
\be\label{tetId2}
e = -\f1{4!}\eps_{IJKL}~\tl\eps^{\m\n\r\s} ~e_\m^I e_\n^J e_\r^K e_\s^L.
\ee
Accordingly,
\begin{subequations}
\begin{align}
\label{ee}
 4e^{[\m}_Ie^{\n]}_J &= - \eps_{IJKL}\eps^{\m\n\r\s} e_\r^Ke_\s^L, \\ \label{eee}
6e^{[\m}_Ie^{\n}_Je^{\r]}_K &= - \eps_{IJKL}\eps^{\m\n\r\s} e_\s^L,
\end{align}
\end{subequations}
which are used in the main text.

For the Hodge dual operator, $\star: \L^p\mapsto\L^{n-p}$ satisfies
\be
\star^2 \om^{(p)}= -(-1)^{p(n-p)}\om^{(p)}, \qquad \om^{(p)}\w\star\th^{(q)} = \om^{(p)}\lrcorner \th_{(q)} \sqrt{-g} d^nx.
\ee
In components, 
\begin{subequations}
\begin{align}
(\star \om^{(p)})^{\m_1..\m_{n-p}} &:= \f1{ p!} \om^{(p)}_{\a_1..\a_p} \eps^{\a_1..\a_p\m_{1}..\m_{n-p}},\\
\om^{(p)}_{\a_1..\a_p} &:=- \f{1}{(n-p)!}\eps_{\a_1..\a_p\m_1..\m_{n-p}} (\star \om^{(p)})^{\m_1..\m_{n-p}}, \\
(\star \om^{(p)})_{\m_1..\m_{n-p}} &:= \f{1}{p!} \om^{(p)}{}^{\a_1..\a_p} \eps_{\a_1..\a_p\m_{1}..\m_{n-p}},  \\
\om^{\a_1..\a_{p}} &:= - \f1{(n-p)!} \eps^{\a_1..\a_p\m_{1}..\m_{n-p}}  (\star \om^{(p)})_{\m_1..\m_{n-p}}.
\end{align}
\end{subequations}

\section{Dressing 2-form}\label{AppT1}

To make some manipulations with the dressing 2-form of \cite{DePaoli:2018erh} more manifest, we report here some useful explicit formulas.
First, we have
\begin{subequations} \label{alpha}
\begin{align}
\star (e_I\w \d e^I) &= -\f12\eps_{IJKL}~ e^I\w e^J\, (e^{K\a}\d e_{\a}^L) = \eps_{\m\n}^{\;\;\;\;\r\s}e_{I\r}\d e_{\s}^I \, dx^\m\w dx^\n, \\
e_I\w \d e^I &= -\eta_{IJKL} ~e^I\w e^J\, (e^{K\a}\d e_{\a}^L) = e_{I[\m}\d e_{\n]}^I \, dx^\m\w dx^\n.
\end{align}
\end{subequations}
Second, the explicit expression of the Hodge dual of the exact 3-form $d\a(\d)$ is, in the general case including the Barbero-Immirzi term, 
\begin{align}
(\star d\a)^\m &= \f1{3!} (d\a)_{\n\a\b} \eps^{\n\a\b\m} = \f1{2} \p_\n\a_{\a\b} \eps^{\n\a\b\m} 
= \f12\p_\n \left(\eps_{\a\b}^{\;\;\;\;\r\s}e_{I\r}\d e_{\s}^I +\f2\g  e_{I[\a}\d e_{\b]}^I\right) \eps^{\n\a\b\m} \nn \\
&= \f12\p_\n \left(e e_{I\r}\d e_{\s}^I - \f1{2\g}  e_{I\g}\d e_{\d}^I \tl\eps^{\;\;\;\;\g\d}_{\r\s}\right)\tl\eps_{\a\b}^{\;\;\;\;\r\s}\eps^{\n\a\b\m}\nn\\
&= -2\p_\n \left(e e^{I[\n} \d e^{\m]I} + \f1{2\g} \tl\eps^{\m\n\g\d} e_{I\g}\d e_{\d}^I \right),
\end{align}
where in the final step we used
\be
e^\n_I g^{\m\s}\d e_{\s}^I = - e^\m_I \d e^{I\n}.
\ee
Finally when torsion is present, we have the following additional relations
\begin{subequations}
\begin{align}
& \p_\s(2e e^{[\s}_I  \d e^{\m]I})  = \na_{\s}\left(2e^{[\s}_I  \d e^{\m]I}\right) - T^\m{}_{\n\r}e^\n_I\d e^{I\r},\\
& \p_\s (\tl\eps^{\m\n\r\s}e_{I\n}\d e^I_\r) = \eps^{\m\n\r\s} \Big[\na_\s(e_{I\n}\d e^I_\r) 
- \f12 T^\l{}_{\s\n}(\d e_{I\l}e^I_\r - e_{I\l}\d e_\r^I) \Big],
\end{align}
\end{subequations}
in agreement with \Ref{divtor}.

\newpage
\bibliography{bibliosimo}

\end{document}